\documentclass[twocolumn,superscriptaddress,10pt,showpacs,aps,prl,floatfix,notitlepage]{revtex4-2}

\usepackage[latin9]{inputenc}
\usepackage{textcomp}
\usepackage{amsmath}
\usepackage{amssymb}
\usepackage{graphicx}
\usepackage{esint}
\usepackage{bm}
\usepackage{comment}
\usepackage{natbib}
\usepackage{hyperref}
\usepackage{mathtools}

\usepackage{cancel}

\hypersetup{colorlinks,%
citecolor=blue,%
filecolor=black,%
linkcolor=blue,%
urlcolor=blue,%
}
\begin{document}

\title{Giant optomechanical coupling and dephasing protection with cavity exciton-polaritons}

\author{P. Sesin}
\affiliation{Centro At\'omico Bariloche and Instituto Balseiro, Comisi\'on Nacional de Energ\'ia At\'omica (CNEA) - Universidad Nacional de Cuyo (UNCUYO), 8400 Bariloche, Argentina.}
\affiliation{Instituto de Nanociencia y Nanotecnolog\'ia (INN-Bariloche), Consejo Nacional de Investigaciones Cient\'ificas y T\'ecnicas (CONICET), Argentina.}
\author{A. S. Kuznetsov}
\affiliation{Paul-Drude-Institut f\"ur Festk\"orperelektronik, Leibniz-Institut im Forschungsverbund Berlin e.V., Hausvogteiplatz 5-7, 10117 Berlin, Germany.}
\author{G. Rozas}
\affiliation{Centro At\'omico Bariloche and Instituto Balseiro, Comisi\'on Nacional de Energ\'ia At\'omica (CNEA) - Universidad Nacional de Cuyo (UNCUYO), 8400 Bariloche, Argentina.}
\affiliation{Instituto de Nanociencia y Nanotecnolog\'ia (INN-Bariloche), Consejo Nacional de Investigaciones Cient\'ificas y T\'ecnicas (CONICET), Argentina.}
\author{S. Anguiano}
\affiliation{Centro At\'omico Bariloche and Instituto Balseiro, Comisi\'on Nacional de Energ\'ia At\'omica (CNEA) - Universidad Nacional de Cuyo (UNCUYO), 8400 Bariloche, Argentina.}
\affiliation{Instituto de Nanociencia y Nanotecnolog\'ia (INN-Bariloche), Consejo Nacional de Investigaciones Cient\'ificas y T\'ecnicas (CONICET), Argentina.}
\author{A. E. Bruchhausen}
\affiliation{Centro At\'omico Bariloche and Instituto Balseiro, Comisi\'on Nacional de Energ\'ia At\'omica (CNEA) - Universidad Nacional de Cuyo (UNCUYO), 8400 Bariloche, Argentina.}
\affiliation{Instituto de Nanociencia y Nanotecnolog\'ia (INN-Bariloche), Consejo Nacional de Investigaciones Cient\'ificas y T\'ecnicas (CONICET), Argentina.}
\author{A. Lema\^itre}
\affiliation{Universit\'e Paris-Saclay, CNRS, Centre de Nanosciences et de Nanotechnologies, 91120, Palaiseau, France.}
\author{K. Biermann}
\affiliation{Paul-Drude-Institut f\"ur Festk\"orperelektronik, Leibniz-Institut im Forschungsverbund Berlin e.V., Hausvogteiplatz 5-7, 10117 Berlin, Germany.}
\author{P. V. Santos}
\affiliation{Paul-Drude-Institut f\"ur Festk\"orperelektronik, Leibniz-Institut im Forschungsverbund Berlin e.V., Hausvogteiplatz 5-7, 10117 Berlin, Germany.}
\author{A. Fainstein}
\email[Corresponding author e-mail: ]{afains@cab.cnea.gov.ar}
\affiliation{Centro At\'omico Bariloche and Instituto Balseiro, Comisi\'on Nacional de Energ\'ia At\'omica (CNEA) - Universidad Nacional de Cuyo (UNCUYO), 8400 Bariloche, Argentina.}
\affiliation{Instituto de Nanociencia y Nanotecnolog\'ia (INN-Bariloche), Consejo Nacional de Investigaciones Cient\'ificas y T\'ecnicas (CONICET), Argentina.}

\begin{abstract}
Electronic resonances can significantly enhance the photon-phonon coupling in cavity optomechanics, but are normally avoided due to absorption losses and dephasing by inhomogeneous broadening. We experimentally demonstrate that exciton-polaritons in semiconductor microcavities enable GHz optomechanics with single-particle resonant couplings reaching record values in the 10s of MHz range. Moreover, this resonant enhancement is protected from inhomogeneous broadening by the Rabi gap. Single-polariton non-linearities and the optomechanical strong-coupling regime become accessible in this platform.
\end{abstract}
\maketitle

%\section{Introduction}

%\paragraph*{\textbf{Introduction.}} 

The search for strong optical forces (i.e., large optomechanical coupling factor $g_0$) is relevant to attain single-photon optomechanical non-linearities (as e.g., optomechanical cooling and self-oscillation)~\cite{RMP} and to access the optomechanical strong and ultra-strong coupling regimes~\cite{RMP,Forn-Diaz2019,Frisk2019,Hughes2021}. 
%These are relevant for the efficient transduction of signals between the optical and acoustic domains at the single-particle level, and for coherent state preparation~\cite{LZSM}. 
Optomechanical non-linearities at the single-photon level are accesible when the optomechanical cooperativity $C_0=4g_0^2/\kappa \Gamma>1$, with $\kappa$ and $\Gamma$ the optical and mechanical dissipation rates, respectively~\cite{RMP}. For the multiple-photon case, the optomechanical coupling becomes amplified as $g_\mathrm{eff}=g_\mathrm{0}  \sqrt{N_\mathrm{p}}$~\cite{RMP}, with $N_\mathrm{p}$ the photon number. The optomechanical strong-coupling regime, attained when $g_\mathrm{eff} > \kappa,\Gamma$~\cite{RMP}, thus requires large optomechanical couplings and small photon and phonon decay rates, but can also be enforced by strong pumping. 
Typically, cavity-optomechanical systems rely on radiation-pressure (RP) forces, i.e. the direct transfer of impulse of a photon when reflecting on a surface~\cite{Cohadon1999}. This mechanism is relatively weak~\cite{Baker2014}, and thus optomechanical non-linearities demand ultra-long photon and phonon lifetimes~\cite{Rossi2018,Ren2020}.  Semiconductor materials provide an alternative strategy through the access to exciton-mediated electrostrictive forces (based on deformation potential interaction, DP), which are enhanced at electronic resonances~\cite{Fainstein2013}. The question is whether this mechanism can attain single particle optomechanical cooperativity $C_0>1$, and if decoherence stemming from resonance-related inhomogeneous broadening and absorption can be circumvented. 
%We address these questions both experimentally and theoretically, and answer them positively.

Research on electrostrictive forces in condensed-matter cavity optomechanics typically avoids optical absorption by working away from electronic resonances~\cite{Rakich2010,Rakich2012,Allain2021}.  An unwanted consequence is that electrostriction decreases and becomes similar in magnitude to radiation pressure~\cite{Baker2014}. Resonance can amplify light-sound coupling through DP  interaction by several orders of magnitude, as previously demonstrated with GaAs multiple quantum wells~\cite{Jusserand2015,Scherbakov2022}. Our aim is to translate this to the context of optical resonators.
Microcavities with embedded GaAs quantum wells (QWs) at resonance in the strong-coupling regime are governed by polaritons, quasi-particles that share the properties of the constituent photons and excitons in ratios that can be tuned through their energy difference~\cite{CarusottoRMP2013}. 
Polariton optomechanics has been theoretically considered with prospects of new phenomena at the quantum limit~\cite{Kyriienko2014,Restrepo2014,Restrepo2017,Vyatkin2021}.  Optomechanical non-linearities including polariton-driven parametric self-oscillation~\cite{Chafatinos2020,Reynoso2022} and asynchronous energy locking~\cite{Chafatinos2022} have been experimentally demonstrated for many-particle condensates  confined in arrays of traps.  Recently, it has been theoretically argued that in micrometer-size cavities with QWs non-linearities can enhance the optomechanical coupling so that $C_0>1$ becomes accesible~\cite{Zambon2022}. The strong-coupling regime in quantum electrodynamics has also been proposed as a means for protecting ensambles of states against decoherence induced by their inhomogeneous broadening~\cite{Diniz2011,Putz2014}.  To the best of our knowledge, however, the resonant enhancement of the optomechanical coupling $g_0$ in the polariton regime has not been experimentally determined, nor has the effect of dephasing been clarified. In this work we experimentally show, through resonant experiments in the strong-coupling regime, that giant values of $g_0$ can be attained. Furthermore,  we demonstrate that the optomechanical resonant coupling is protected from dephasing induced by the inhomogeneous distribution of resonant states because of the opening of the polariton Rabi gap. For this purpose we present two complementary experimental approaches, namely photoluminescence in the presence of piezoelectrically injected resonant bulk acoustic waves (BAWs), and Brillouin scattering as a function of exciton-photon detuning and temperature.

%\paragraph*{\textbf{Microcavity samples and experimental set-up.}} 
Optical cavities with distributed Bragg reflectors also confine acoustic phonons~\cite{Trigo2002,Fainstein2013}. We study two different microcavity structures, specifically designed for polariton modulation with electrically generated bulk acoustic waves (sample A), or for Brillouin scattering experiments (sample B). Both are planar structures grown by molecular beam epitaxy, with a thickness taper to allow the variation of the cavity-exciton detuning $\delta$=$E_c-E_x$, where $E_c (E_x)$ is the cavity (exciton) energy.

Sample A is an (Al,Ga)As microcavity with a spacer layer embedding six 15-nm-thick GaAs QWs~\cite{SM,Kuznetsov2021} and designed to confine $\sim$1.53 eV photons and $\omega_m/2\pi \sim 7$~GHz phonons. 
%The spacer region acts as an optical $5 \lambda/2$ and as an acoustic $5/3 \lambda/2$ microcavity spacer containing two 15-nm-thick In$_{0.04}$Ga$_{0.96}$As QWs separated by a 5-nm-thick GaAs barrier. 
The QWs are positioned inside the spacer at antinodes of the optical field and close to an antinode of the acoustic strain field to enhance the DP coupling~\cite{Kuznetsov2021}. The QWs display heavy-hole (hh) and light-hole (lh) excitonic transitions separated by $\sim 5$~meV, which strongly couple to the confined photon state, leading to three polariton branches with Rabi gaps of $\sim 7$ and $\sim 5$~meV, respectively~\cite{SM}. Monochromatic phonons with tunable amplitudes and frequencies are injected into the microcavity using radiofrequency-driven BAW ZnO resonators, fabricated on the surface of the structure.  These resonators also act as phonon detectors, enabling acoustic-echo spectroscopy of the microcavity~\cite{Machado2019}. A ring-shaped geometry with apertures in the bottom and top contacts allows optical access to the microcavity~\cite{SM}.

Sample B  is a planar semiconductor microcavity with a thick $9\lambda/2$ spacer constituted by a 41.5-periods GaAs/AlAs 17.1/7.5~nm multiple quantum well (MQW)~\cite{SM,Jusserand2015}. 
%The MQW is identical to the one studied in Ref.~\cite{Jusserand2015} which, however, had no optical confinement. 
%Each period consists of a 17.1~nm GaAs quantum well and a 7.5~nm AlAs barrier.  
A MQW is also a periodically modulated acoustic structure, so its acoustic properties can be described in the band-folding scheme~\cite{JusserandLSS}. We perform resonant Brillouin backscattering experiments between 30 and 80~K, coupling to a $\omega_m/2\pi \sim180$~GHz zero-group-velocity (ZGV) mode that results from this folding. 
Brillouin scattering is performed in double optical resonance (DOR)~\cite{Fainstein1995,Rozas2014}, i.e. the angles and energies are set so that both incoming and outgoing photons are tuned to polariton resonances of the lower branch.

%\paragraph*{\textbf{Experimental determination of $g_\mathrm{om}/2\pi$.}} 
%%%%%%%%%%%%%%%%%%%%%%%%%%%%%%%%%%%%%%%%%%%%%
\begin{figure}[!hht]
 \begin{center}
\includegraphics*[keepaspectratio=true, clip=true, trim = 0mm 0mm 0mm 0mm, angle=0, width=1\columnwidth]{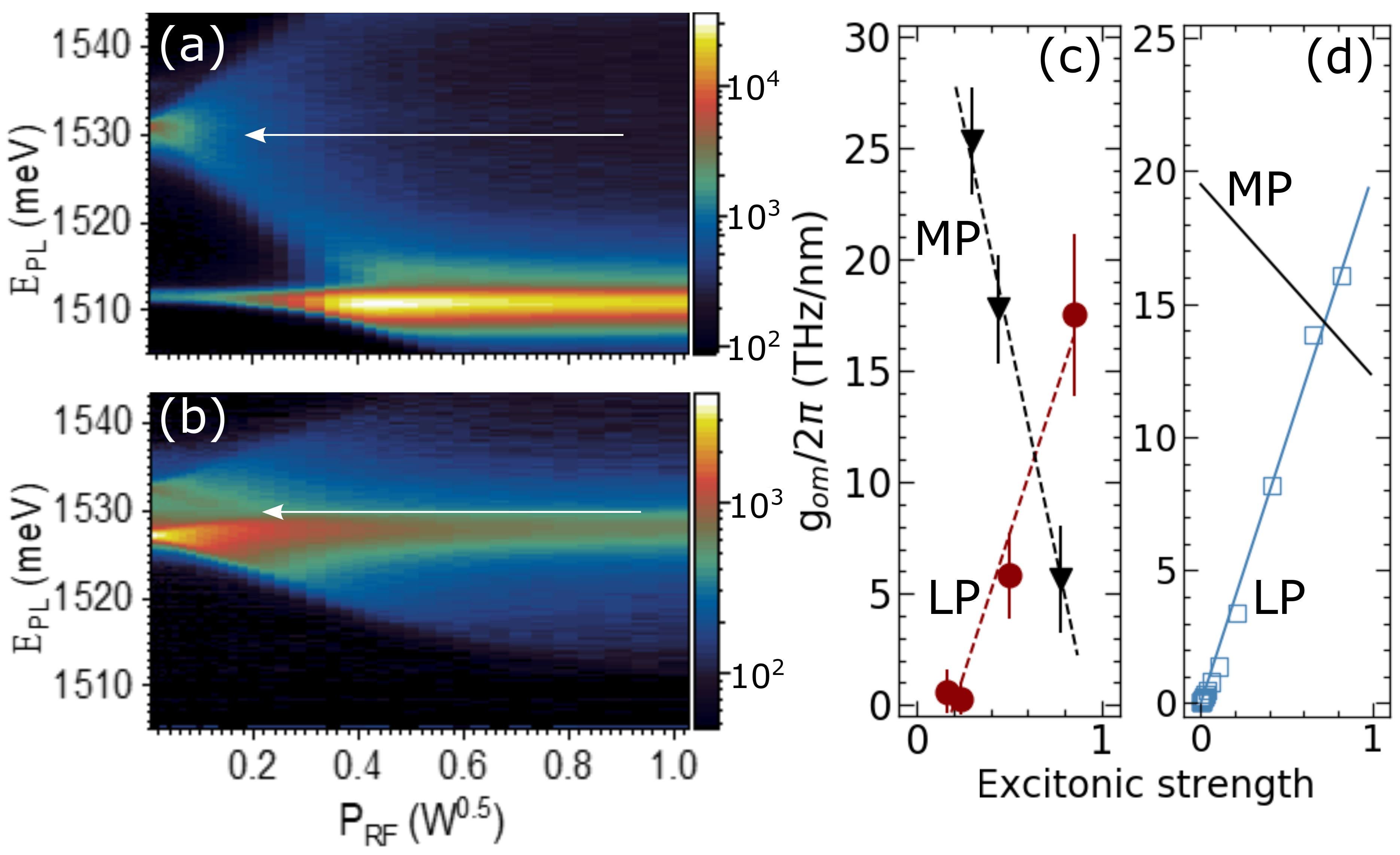}
\end{center}
%\vspace{-0.8 cm}
%    \hspace{-1.6 cm}
\caption{
%\textbf{Decay energy for Lower Polariton at different temperatures.}
Panels (a) and (b) present the time-integrated polariton emission of sample A for non-resonant laser excitation as a function of radio-frequency power applied to the piezoelectric transducer. Panel (a) correspond to a negative detuning of about -17.5~meV, so that the lower polariton centered at $\sim 1.511$~eV is mostly photonic in character, while the middle polariton at $\sim 1.531$~eV is mostly heavy-hole excitonic (the heavy-hole exciton energy $E_x$$\sim$1.530~eV is indicated with an horizontal white arrow). Panel (b) is the case for zero detuning. 
Panels (c) and (d) display the experimental and theoretical optomechanical coupling factor $g_{om}/2\pi$, respectively, as a function of the exciton strength of the corresponding polariton branch (heavy-hole strength for the lower polariton LP, and light-hole strength for the middle polariton MP). In (c), solid circles (triangles) correspond to the LP (MP). Dashed lines are guides to the eye.  In (d), squares correspond to the theoretical model presented in the text, with no adjustable parameters. The continuous lines indicate the excitonic Hopfield coefficient scaled to fit $g_{om}/2\pi$ at the largest detunings both for the LP and the MP branches (see text for details).}
\label{Fig01}
\end{figure}
%%%%%%%%%%%%%%%%%%%%%%%%%%%%%%%%%%%%%%%%%%%%%
Figures~\ref{Fig01}(a)  and \ref{Fig01}(b) present color maps of PL spectra  (10~K)  obtained on sample A as a function of the BAW amplitude for negative and zero photon-exciton detuning $\delta$, respectively. These were obtained under low non-resonant optical excitation powers (below the polariton condensation threshold), and by resonantly driving the cavity with the $\sim 7$~GHz acoustic mode of the resonator~\cite{SM}. For the negative detuning in Fig.~\ref{Fig01}(a), the heavy-hole excitonic-like resonance can be identified at $\sim$1.531~eV and the cavity-like mode at $\sim$1.511~eV.  It is apparent, and particularly striking in this case where the polariton modes resemble their bare constituents, that the BAWs mostly affect the exciton's energy (through deformation potential interaction), and only very little the photon cavity mode (reflecting the weak radiation-pressure contribution). For the zero-detuning case in Fig.~\ref{Fig01}(b),  the lower and middle polariton modes can be observed. Both split symmetrically around the exciton energy by the Rabi gap $\sim$7~meV, and become modulated in similar amounts by the externally applied BAWs.
% (at $\delta=0$ the two polariton states are approximately half exciton and half photon).

%$g_\mathrm{om}/2\pi$ is precisely the quantity measured using electrically generated BAWs, and to be presented in the next section. 
We derive the mechanically induced LP and MP energy modulation $\Delta E_\mathrm{LP(MP)}$ from the PL spread of the respective mode in Figs.~\ref{Fig01}(a) and (b). The strain $s$ at the QWs for a given radio-frequency power $P_{RF}$ is calculated from the energy injected by the piezoelectric BAW resonators, and using a transfer-matrix model for the spatial distribution of the acoustic cavity mode~\cite{SM}. From it we derive the displacement $\Delta u = s/q$, with $q$ the phonon wavenumber. We thus obtain $g_\mathrm{om}/2\pi = \Delta E_\mathrm{LP(MP)}/\Delta u$, i.e., the change of polariton (LP or MP) energy per unit displacement, which is shown in Fig.~\ref{Fig01}(c) with the dark red circles and black triangles for the LP and MP modes, respectively, as a function of the excitonic strength of the corresponding polariton mode. The excitonic strength refers to the heavy-hole component for the LP, and to the light-hole for the MP, both which vary from 0 to 1 when going from very negative to very positive detuning. At very negative detuning the LP mode is mostly photonic, while the MP is fully heavy-hole excitonic.

$g_\mathrm{om}/2\pi$ can be related to the single-particle linear optomechanical coupling $g_\mathrm{0}=g_\mathrm{om} x_\mathrm{zpf}$~\cite{RMP}. $x_\mathrm{zpf}$ is the displacement due to zero-point fluctuations,  and depends on the precise 3D geometry of the microstructured optomechanical resonator~\cite{Zambon2022,Anguiano2017,Lamberti2017}. 
The maximum magnitude measured at large heavy-hole excitonic fractions is $g_{om}/2\pi \sim 15-20$~THz/nm. To grasp the implications of this experimentally determined value, one can consider a pillar of circular shape and $1.2~\mu$m-diameter for which  we estimate $x_\mathrm{zpf} \sim 1\mathrm{pm}$~\cite{SM,Villafane2018}. Using the measured $g_{om}/2\pi$ the corresponding single-polariton optomechanical factor is $g_{0}/2\pi \sim 15-20$~MHz, which can be considered record-high in the cavity-optomechanics literature. The origin of such giant optomechanical coupling, and its strong detuning dependence, will be theoretically addressed next.

%\section{Experimental details}
%%%%%%%%%%%%%%%%%%%%%%%%%%%%%%%%%%%%%%%%%%%%%
\begin{figure}[!hht]
 \begin{center}
\includegraphics*[keepaspectratio=true, clip=true, trim = 0mm 0mm 0mm 0mm, angle=0, width=0.9\columnwidth]{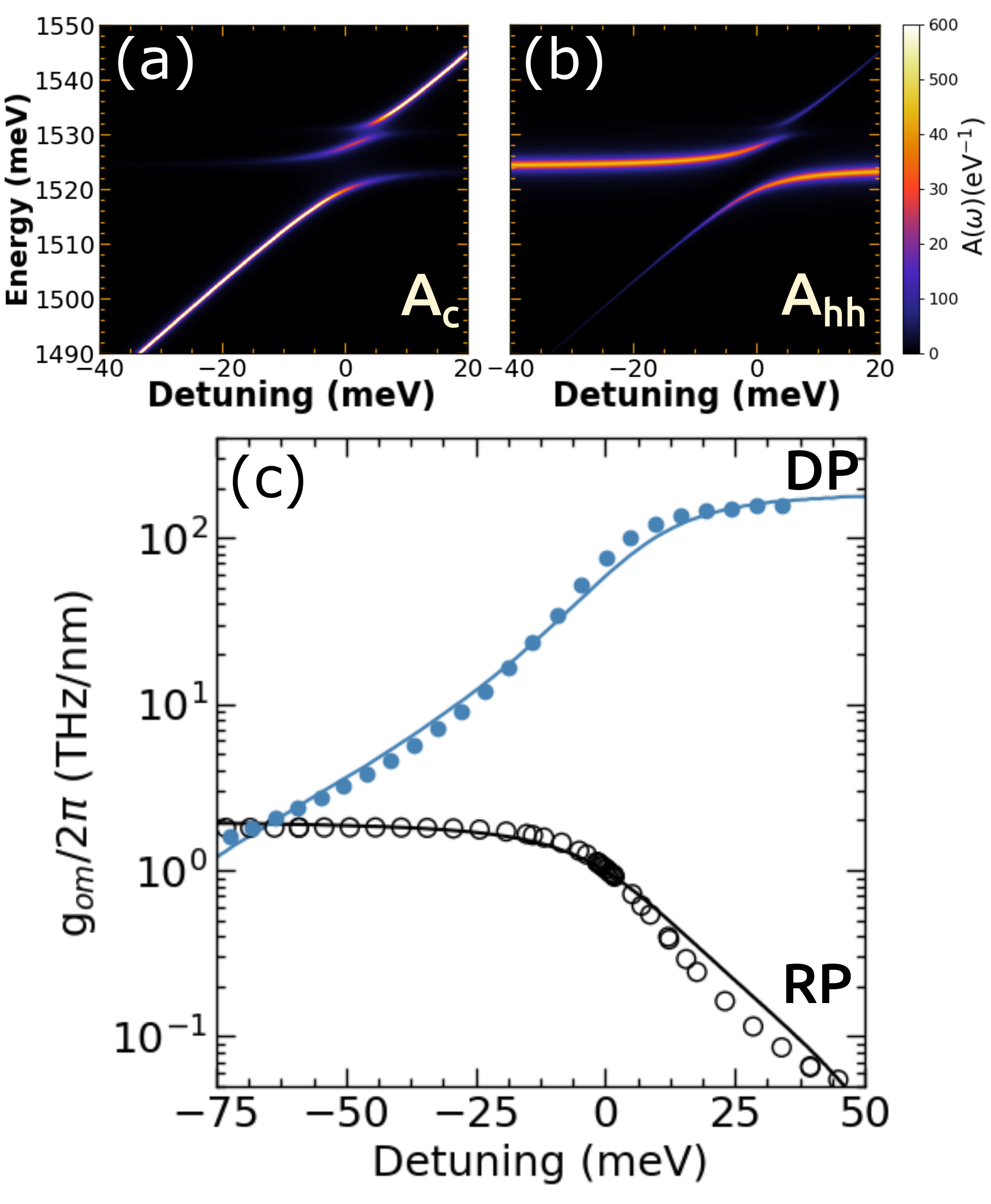}
\end{center}
%\vspace{-0.8 cm}
%    \hspace{-1.6 cm}
\caption{
%\textbf{Cavity Polaritons photoluminescence spectral maps vs detuning,  at 293$\,$K (a), 78$\,$K (b), and 30$\,$K (c).}
Panels (a) and (b) are color-intensity maps of the photon and heavy-hole spectral weights, respectively, calculated for the polariton branches of sample B. The symbols in panel (c) are the calculated optomechanical coupling factor $g_{om}/2\pi$, due to radiation-pressure (RP, open circles) and deformation-potential (DP, solid circles) interaction. The solid curves, superimposed on the RP and DP contributions, correspond to the scaled excitonic and photonic Hopfield coefficients, respectively. 
}
\label{Fig02}
\end{figure}
%%%%%%%%%%%%%%%%%%%%%%%%%%%%%%%%%%%%%%%%%%%%%

%\paragraph*{\textbf{Theoretical considerations.}} 
To introduce the model used to calculate the optomechanical coupling, we consider sample B. The QWs in this sample also display two energetically close heavy-hole (hh) and light-hole (lh) excitonic transitions, which strongly couple to the confined photon state, leading to three polariton branches~\cite{SM}. The low-temperature (30~K) detuning dependence of the  photon ($A_c$) and hh-exciton ($A_{hh}$) spectral densities of these polariton branches, derived from PL experiments~\cite{SM}, are shown in Figs.~\ref{Fig02}(a) and \ref{Fig02}(b), respectively.  The integral of these spectral densities over each branch ($S_{c(hh)} = \int_\omega d\omega A_{c(hh)}$) identifies the fraction of each bare state on the coupled modes (i.e., the Hopfield coefficients $S_{c(hh)}$). The hh (lh) Rabi splitting is $\Omega_{hh(lh)}= $~8.7 (4.3)~meV, and is almost temperature independent in the studied range ($T < 150$~K).  We focus our attention here on the LP branch, which, according to Figs.~\ref{Fig02}(a) and \ref{Fig02}(b), evolves from purely photonic to purely hh excitonic when going from negative to positive detuning (the lh contribution to this state is negligible).

$g_\mathrm{om}/2\pi$ (shown in Fig.~\ref{Fig02}c) is calculated for a planar structure as the shift induced by a mechanical perturbation on the absorptance peak associated with the LP, and calculated as  $A=1-R-T$. $R$ and $T$ are the reflectance and transmitance, respectively, obtained with the transfer-matrix method. The excitonic resonance is included in the dielectric function of the QWs  as
%\begin{equation}
%\label{EQ01}
$\varepsilon(\omega) = \varepsilon_\infty + \sum_l \frac{4\pi \beta_l \omega_l^2 }{\omega_l^2 - \omega^2 - i \Gamma_l \omega}$~ \cite{Tredicucci1995,Chen1995},
%\end{equation}
where $\varepsilon_\infty$ is the background permitivity, $\omega_l$ is the l-th excitonic energy (l=hh or lh), and $\Gamma_l$ is the corresponding excitonic linewidth. This model describes well, in a classical way, the avoided crossing between excitons and photons in the cavity~\cite{SM}. The excitons' oscillator strengths $4\pi \beta_l \omega_l^2$ are set to fit the experimentally observed Rabi gaps.  With a similar transfer-matrix method, the acoustic modes can also be derived~\cite{SM}, and, with the obtained spatial distribution, their effect on the absorptance $A$ can be evaluated. The DP contribution corresponds to the effect of strain $s=\partial u/\partial z$ on the excitons' energies, included  as $\tilde{\omega}_{l} = \omega_{l} - \Xi_{l} \times s$, where $\Xi_{hh (lh)}$ is the hh~(lh) deformation potential (DP) coefficient. Radiation pressure (RP) corresponds to the change of polariton energies induced by the movement of the interfaces within the structure due to the displacement $u(z)$. We note that an additional contribution to $g_\mathrm{om}/2\pi$ arises from the non-resonant photoelastic effect, i.e. the strain-induced variation of $\varepsilon_\infty$. Physically, this term emerges from all DP contributions associated to higher-energy, non-resonant gaps. Its detuning dependence follows that of the RP term, with a similar magnitude~\cite{Villafane2018}. For simplicity, in what follows, we will consider this photoelastic contribution included within RP. Standard parameters were used for the calculation of $g_\mathrm{om}/2\pi$ presented in Fig.~\ref{Fig02}(c)~\cite{SM}.

Two results can be highlighted in Fig.~\ref{Fig02}(c). First, except for negative detunings below $\sim-50$~meV, where a crossover is observed, the DP largely surpasses the RP contribution, being approximately two orders of magnitude larger at zero detuning (note the logarithmic scale in Fig.~\ref{Fig02}(c)). Second, the overall detuning dependence of the RP and DP contributions to $g_\mathrm{om}/2\pi$ follow the photonic ($S_c$) and excitonic ($S_{hh}$) Hopfield coefficients, respectively. These coefficients, shown with solid curves in Fig.~\ref{Fig02}(c), are scaled to fit the data at the maximum detunings. This result justifies the {\it ad-hoc} assumption taken in Refs.~\cite{Chafatinos2020,Zambon2022}, which states that $g_{0}=S_c\,g_{0}^{RP} + S_{hh}\,g_{0}^{DP}$.
% where $S_c$ and $S_{hh}$  are the squared spectral density for photons ($\left|A_c(\omega)\right|^2$) and for hh excitons ($\left|A_{hh}(\omega)\right|^2$), respectively.

The same model, when applied to sample A (only based on the specific layered structure and with no fitting parameter), leads for the lower polariton to the results shown with open squares in Fig.~\ref{Fig01}(d). Following the above discussion, we present in the same  figure, with a solid blue line, the excitonic Hopfield coefficient $S_{hh}$ multiplied by a constant (which corresponds to the hh $g_{om}^{DP}$). Based on the same considerations, for the middle polariton mode that varies from being purely hh-like to lh-like (this mode has very little photonic component~\cite{SM}), the corresponding $g_{0}=S_{lh}\,g_{0,lh}^{DP} + S_{hh}\,g_{0,hh}^{DP}$. The two excitonic optomechanical factors $g_{0,lh}$ and $g_{0,hh}$ simply scale as the corresponding deformation potential coefficient $\Xi_{hh (lh)}$ ($\sim 10.5$~eV and $\sim 6.5$~eV for the hh and lh excitons, respectively). As follows from Figs.~\ref{Fig01}(c) and (d), there is an excellent quantitative agreement between experiment and theory, demonstrating our conclusion that the main contribution to $g_{0}$ is the DP interaction, which follows the detuning dependence of the excitonic component of the polariton.

%%%%%%%%%%%%%%%%%%%%%%%%%%%%%%%%%%%%%%%%%%%%%
\begin{figure}[!hht]
 \begin{center}
\includegraphics*[keepaspectratio=true, clip=true, trim = 0mm 0mm 0mm 0mm, angle=0, width=1\columnwidth]{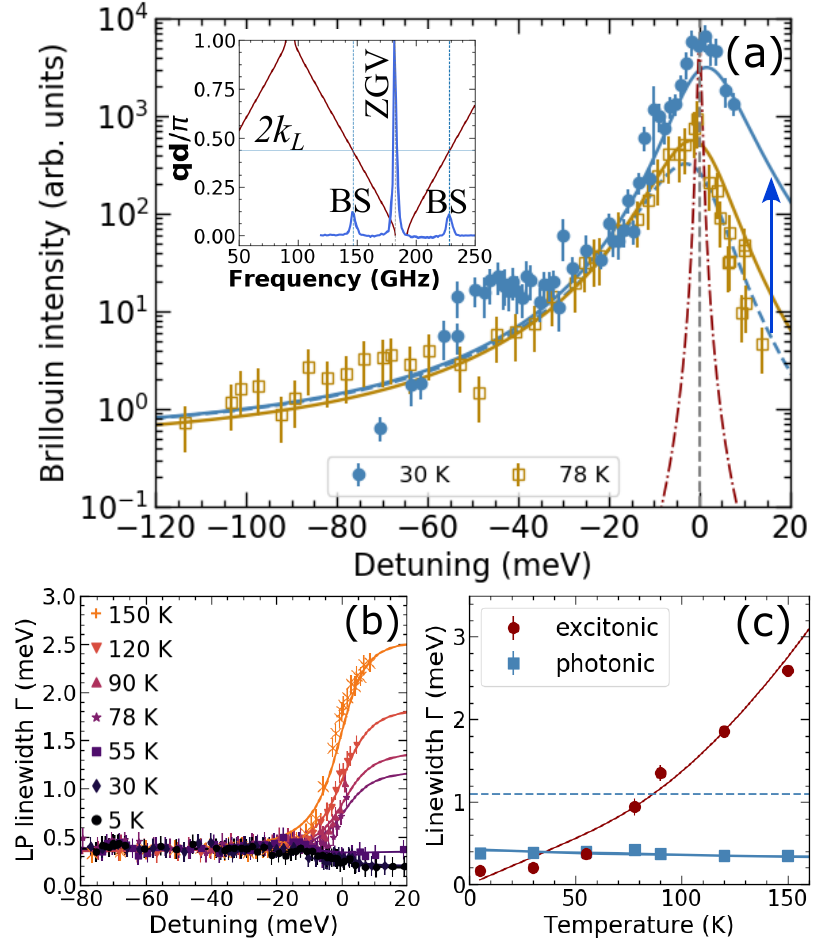}
\end{center}
%\vspace{-0.8 cm}
%    \hspace{-1.6 cm}
\caption{
%\textbf{Brillouin Scattering Intensity vs Polariton detuning}
(a) Brillouin scattering intensity on sample B as a function of detuning for 30 and 78~K. Symbols represent the experiments, and solid curves the model that accounts for both the RP- and DP-mediated optomechanical interactions.  
The inset presents an example of a Brillouin spectrum showing the studied zero-group-velocity mode (ZGV), and the two backscattering peaks (BS). The phonon dispersion in the folded-band scheme is also shown. The horizontal line at $2k_L$ indicates the BS transferred wavenumber. The effect of inhomogeneous broadening on the 30~K case is shown with the dashed steel-blue curve (the blue arrow indicates the dephasing protection by strong coupling). The red dashed-dotted curve is the detuning dependence for an identical, but bare (no cavity confinement), MQW, as extracted from Ref.~\onlinecite{Jusserand2015}.
Panel (b) presents the measured detuning dependence of the LP full linewidth at different temperatures. Panel (c) shows the extracted temperature dependence of the photon and heavy-hole exciton linewidths. The horizontal dashed line corresponds to the exciton inhomogeneous broadening.
Thin solid curves in panels (b) and (c) are fits with the theoretical models explained in the text.  
}
\label{Fig03}
\end{figure}
%%%%%%%%%%%%%%%%%%%%%%%%%%%%%%%%%%%%%%%%%%%%%

%\paragraph*{\textbf{Strong-coupling dephasing protection.}} 
Information on the detuning dependence of the optomechanical coupling factor can also be obtained through the resonant dependence of the Brillouin intensity. This is shown in Fig.~\ref{Fig03}(a) for two temperatures, 30 and 78~K,  for inelastic scattering by the ZGV mode at $\sim$180~GHz in sample B (a typical Brillouin spectrum is presented in the inset of this figure).  Note the four-orders-of-magnitude resonant increase of the scattering efficiency observed at 30~K.  The magnitude of this enhancement is comparable to that observed for a similar MQW \emph{without} photonic confinement~\cite{Jusserand2015}. The detuning dependence is, however, markedly different (the detuning dependence for the bare MQW is displayed in Fig.~\ref{Fig03}(a) with a red dashed-dotted curve, as extracted from Ref.~\onlinecite{Jusserand2015}).  The steepness of the resonance for the bare MQW is determined by the excitonic linewidth ($\sim 70 \mu$eV at 30~K), while for the microcavity with an embedded MQW in Fig.~\ref{Fig03}(a) the detuning dependence is determined by the excitonic Hopfield coefficient (which varies in the same scale as the Rabi splitting $\sim 9$~meV). The decrease of resonant enhancement with temperature, which for 78~K amounts to one order of magnitude, as shown in Fig.~\ref{Fig03}(a), evidences the role of dephasing. We address these effects next.

Figure~\ref{Fig03}(b) presents the detuning dependence of the LP's full linewidth for different temperatures ranging from 5 to 150~K. It is observed that at 30~K the LP's linewidth is almost detuning independent (decreases slightly when going from photon to hh exciton character), while at 78~K a clear broadening occurs as polaritons become more excitonic (reflecting the phonon-induced exciton dephasing). From these data we extract the temperature dependence of the hh exciton ($\Gamma_{hh}$) and cavity ($\Gamma_{c}$) linewidths~\cite{Sermage1996},  which are presented with symbols in Fig.~\ref{Fig03}(c).  $\Gamma_{c}$ turns out to be temperature independent ($\Gamma_c \sim 0.4$~meV), while $\Gamma_{hh}$ monotonically increases with temperature. Its variation can be fitted~\cite{Gammon1995} with a linear contribution of the acoustic phonons thermal population and a second term proportional to the longitudinal optical phonon population $n_{LO}$: $\Gamma_{hh}(T)=0.02$~[meV]$+1.1 \times 10^{-2}$~[meV/K]$T$+16.6[meV]$n_{LO}$ (shown with the solid thin curve in Fig.~\ref{Fig03}(c))~\cite{Jusserand2015,Gammon1995}. The zero-temperature value $\Gamma_{hh}(0)=0.02$~meV corresponds to an homogeneous linewidth~\cite{Houdre1996}, which contrasts with the much larger measured inhomogeneous spread of exciton states $\Gamma^{inh}_{hh} \sim 1.1$~meV (indicated with an horizontal dashed line in Fig.~\ref{Fig03}(c)). In fact, it is quite notable in Fig.~\ref{Fig03}(b) that at low temperatures the LP linewidth, even at the largest positive detuning, does not reflect the inhomogeneous broadening of the excitonic states.

Polariton-mediated Brillouin scattering can be described as a three-step process~\cite{Rozas2014}: (1) conversion of the impinging photon of frequency $\omega$ to an incoming polariton, followed by (2) the scattering of this polariton into another polariton of the same branch but different in-plane wavevector and frequency $\omega'=\omega-\omega_m$, accompanied by the emission of a ZGV acoustic phonon, and (3) the conversion of this latter polariton into an external photon, which is finally detected.  This sequential process can be modeled as
$I_B \propto S_c(g_{0}^{RP}S_c+ g_{0,hh}^{DP} S_{hh})^2S_{c}\Gamma_{LP}^{-2}$, where it is assumed that the phonon frequency is small so that the Hopfield coefficients corresponding to incoming and scattered polaritons are the same. The lifetime of these two intervening polariton states is included through $\Gamma_{LP}^{-2}$~\cite{SM,Rozas2014}. Note that in this expression for $I_B$ it is assumed that, for the LP branch, $S_{lh} \sim 0$. Theoretical results are shown with solid curves in Fig.~\ref{Fig03}(a). 
Because of the intrinsic uncertainty on the determination of absolute Brillouin cross sections, the experimental curve at 30~K has been scaled to coincide with the theoretical maximum, this being the only adjustable parameter which valid for both temperatures.  Disregarding this unavoidable scaling, the agreement between theory and experiments is noteworthy. The quenching of the resonant enhancement with temperature can be understood as being due to the broadening of the polariton line, which follows the detuning dependence of the excitonic component of the polariton. The dashed steel-blue curve in Fig.~\ref{Fig03}(a) is the calculation assuming that at 30K the hh exciton is also affected by its inhomogeneous broadening. It is clear that the opening of the Rabi gap in the strong-coupling regime protects polaritons against such dephasing,~\cite{Houdre1996} remaining only the phonon-dependent homogeneous linewidth as the limitation factor for the fully resonant, exciton-mediated, DP optomechanical interaction.

%\paragraph*{\textbf{Discussion and Conclusions.}} 
To conclude, we have reported a direct experimental determination of the resonant enhancement of the optomechanical coupling factor in DBR-based microcavities with embedded QWs in the polariton strong-coupling regime. 
Due to the QW's excitonic resonance, the optomechanical coupling is dominated by the deformation-potential interaction, becoming more than two orders of magnitude larger than that due to radiation pressure. 
A polariton-induced dephasing protection occurs because inhomogeneous broadening is avoided due to the opening of the Rabi gap~\cite{Houdre1996}. Polariton strong coupling is thus demonstrated in the context of cavity optomechanics as a path for protection against decoherence~\cite{Diniz2011,Putz2014}.  

Based on the enhanced optomechanical coupling demonstrated in this work, we estimate that optomechanical nonlinearities ($C_0 \sim 1$) can be attained with pillar cavities with a diameter of $\sim 1.2 \mu$m (for which we estimate $g_0/2\pi\sim$15-20~MHz), and with experimentally feasible optical and mechanical Q factors $\sim 5 \times 10^4$. Polariton condensates are also interesting due to the large coherence time of their collective phase (as compared to the decay time of individual polaritons). Based on the coherence time measured in our devices (in the range of 1--2~ns), we conclude that as few as $2 \times 10^3$ polaritons in the condensate are required to attain the optomechanical strong-coupling regime ($g_\mathrm{eff} > \kappa_{LP},\Gamma_m$). 
%Our results close the gap toward future applications of DBR based microcavities with embedded QWs for the efficient and coherent transduction of signals between the optical and acoustic domains, and for the use of mechanical waves for polariton state preparation~\cite{LZSM}. 

%\section{Conclusions}

\begin{acknowledgments}
We acknowledge financial support from the ANPCyT (Argentina) under grant PICT-2018-03255, from German DFG (grant 359162958) and QuantERA grant Interpol (EU-BMBF (Germany) grant nr. 13N14783), and from the French RENATECH network. 
AF acknowledges support from the Alexander von Humboldt Foundation. We also acknowledge the technical support from R. Baumann, S. Rauwerdink, and A. Tahraoui with the sample fabrication process.
\end{acknowledgments}

%\bibliography{citations} 
     % American Physical Society (APS) style, author-year citations
  
          % name your BibTeX data base

%%%%%%%%%%%%%%%%%%%%%%%%%%%%%%%%%%%%%%%%%%%%%%%%%%%%%%%%%%%%%%%%%%%%%%%
%%%%%%%%%%%%%%%%%%%%%%%%%%%%%%%%%%%%%%%%%%%%%%%%%%%%%%%%%%%%%%%%%%%%%%%
%        Beginning of SM
%%%%%%%%%%%%%%%%%%%%%%%%%%%%%%%%%%%%%%%%%%%%%%%%%%%%%%%%%%%%%%%%%%%%%%%
%%%%%%%%%%%%%%%%%%%%%%%%%%%%%%%%%%%%%%%%%%%%%%%%%%%%%%%%%%%%%%%%%%%%%%%

\onecolumngrid

\pagebreak

%\clearpage
\setcounter{section}{0}
\setcounter{page}{1}
\setcounter{figure}{0}

\renewcommand{\figurename}{Figure \!\!}
\renewcommand{\thefigure}{\textsf{S}\arabic{figure}}

\renewcommand{\thesection}{\textsf{S}\,\Roman{section}}

\renewcommand{\theequation}{\textsf{S}\arabic{equation}}

\begin{center}
\textbf{\large Supplementary Material for:\\ Giant optomechanical coupling and dephasing protection with cavity exciton-polaritons}
\end{center}

This supplementary material provides descriptions and details of the experimentally studied samples A and B, setups, and theoretical aspects presented in this letter. It is organized as follows: the studied samples are presented in Section \ref{secSampledetails}. The experimental setups used are described in Section \ref{secExperimentalDetails}.  The theoretical concepts for the cavity polaritons photoluminescence and Brillouin scattering spectroscopy, the optomechanical coupling calculations, and the estimation of the bulk-acoustic-wave electrically injected strain, are discussed in Section \ref{secTheory}. Section \ref{sec: Raman spectra in the cavity - phonon peaks} presents a calculation of the phonon modes and Brillouin spectra of sample B. Finally, we present in Section \ref{g0} the obtained values for the optomechanical coupling and discuss the thresholds for the observation of optomechanical non-linearities based on the experimentally determined optomechanical couplings.

\section{Samples' details}\label{secSampledetails}

Two different semiconductor microcavities are studied. \textbf{Sample A} was specifically designed and fabricated to allow the modulation of polaritons by electrically generated bulk acoustic waves. \textbf{Sample B}, on the other hand, was designed for Brillouin scattering experiments. Both samples consist of planar heterostructures grown on a GaAs (001) substrate by molecular beam epitaxy, with a thickness taper to allow the variation of the cavity-exciton detuning from negative to positive values. 

%%%%%%%%%%%%%%%%%%%%%%%%%%%%%%%%%%%%%%%%%%%%%%%%%%%%%%%%%%%%%%%%%%%%%%%%%%%

\begin{figure}[hhh!!!]
\begin{center}
\includegraphics*[keepaspectratio=true, clip=true, trim = 0mm 0mm 0mm 0mm, angle=0, width=0.6\columnwidth]{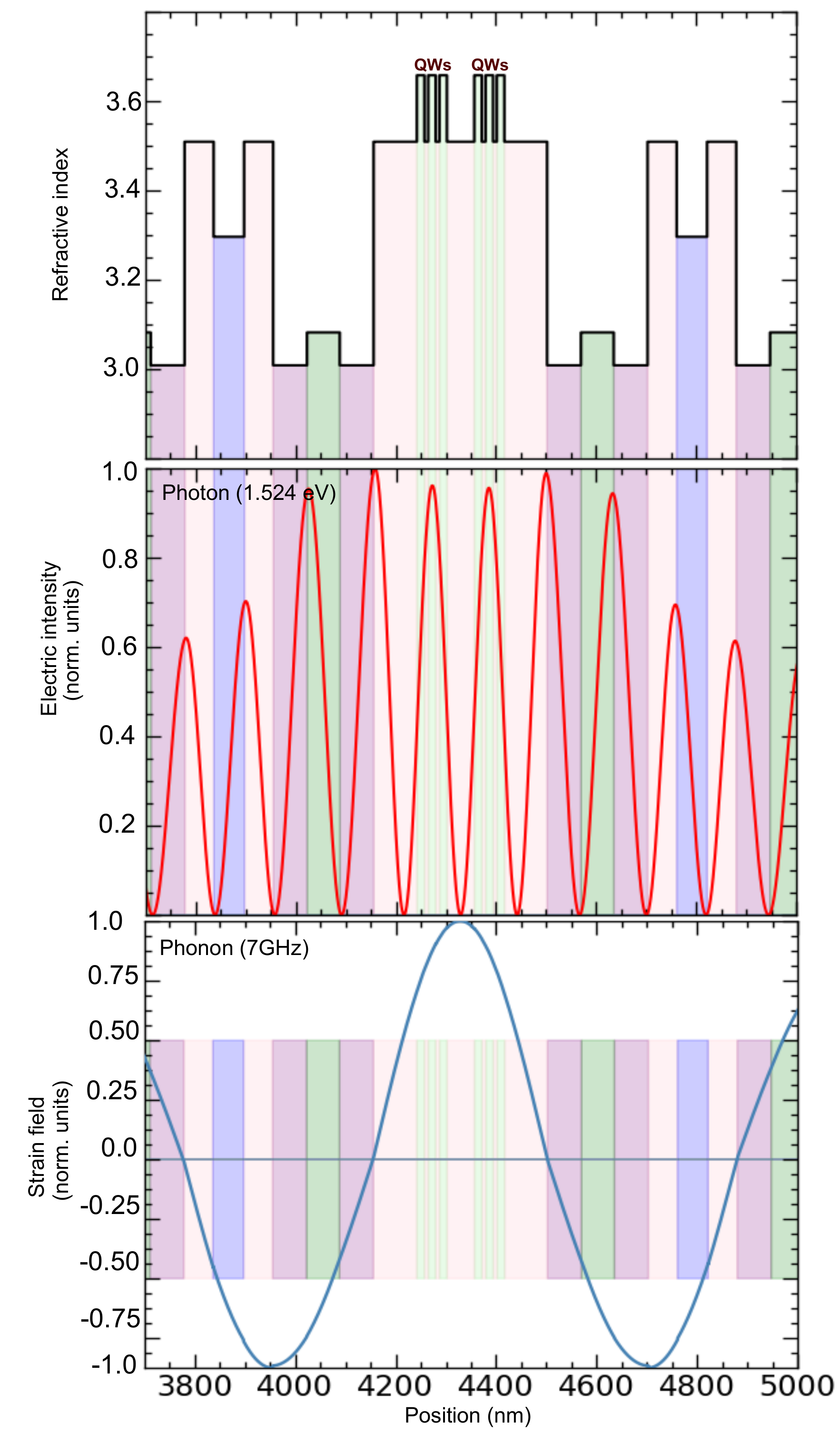}
%\vspace{-0.8 cm}
\caption{(a)  Refractive index profile around the spacer region, plotted along the growth direction of sample A. The position of the embedded quantum wells is green colored. (b) Normalized intensity of the resonant electric field at 1.527 eV. (c) Normalized strain field of the 7 GHz phonon.}
\label{FigS01}
\end{center}
\end{figure}

Sample A is a hybrid acousto-optical (Al,Ga)As microcavity with a spacer layer embedding six GaAs quantum wells (QWs). In Fig. \ref{FigS01}(a) the refractive index profile around this region is plotted. This sample has a particular design that enables the simultaneous confinement of photons  within its spacer layer with a (vacuum) wavelength of  $\sim 820\,$nm [see Fig. \ref{FigS01}(b)] and phonons with a frequency of approximately 7 GHz  [see Fig. \ref{FigS01}(c)]. This sample can also hold a $\sim 20\,$GHz mechanical mode. The spacer region embeds six 15-nm-thick GaAs QWs with 7.5-nm-thick Al$_{0.1}$Ga$_{0.9}$As barrier layers. As schematized in Fig. \ref{FigS01}, these QWs are positioned at a depth corresponding to the antinodes of the squared optical field and close to the antinode of the acoustic strain field in otder to enhance the optomechanical coupling. The spacer is sandwiched between two distributed Bragg reflectors (DBRs). The top (bottom) DBR is constituted by 11(14) periods formed by six $\lambda/4$ layers: Al$_{0.1}$Ga$_{0.9}$As/Al$_{0.3}$Ga$_{0.7}$As/Al$_{0.1}$Ga$_{0.9}$As/Al$_{0.9}$Ga$_{0.1}$As/Al$_{0.7}$Ga$_{0.3}$As/Al$_{0.9}$Ga$_{0.1}$As. Furthermore, this sample allows the injection of phonons into the cavity using bulk-acoustic-wave generators (BAWRs) on the surface, which are driven by an external radio-frequency electric field. BAWRs generate highly monochromatic phonons with tunable amplitudes and frequencies over a range of several GHz, which far exceeds the stopband of the phononic cavity. Also, BAWRs allow phonon detection, thus enabling acoustic-echo spectroscopy, with which the acoustic response of the system is studied \cite{SM_Machado}.

Each BAWR consists of a piezoelectric ZnO film sandwiched between two 50-nm-thick contacts. These contacts are made of Ti/Al/Ti (10/30/10 nm) and are square-shaped, with a central aperture allowing optical access to the microcavity. The deposited ZnO films have their hexagonal $c$ axis oriented normal to the microcavity surface. Their thicknesses are chosen to be $d_{\text{ZnO}}=\lambda_{\textsc{BAW}}/2$, where $\lambda_{\textsc{BAW}}$ is the tunable acoustic wavelength \cite{SM_Machado}. 
A front view of the BAWR coupled to the top of sample A is presented in Fig. \ref{FigS03}. The acoustoelectric response of the microcavity is probed using a vector network analyzer to record its RF-power reflection coefficient.  
%%%%%%%%%%%%%%%%%%%%%%%%%%%%%%%%%%%%%%%%%%%%%%%%%%%%%%%%%%%%%%%%%%%%%%%%%%%

\begin{figure}[hhh!!!]
\begin{center}
\includegraphics*[keepaspectratio=true, clip=true, trim = 0mm 0mm 0mm 0mm, angle=0, width=0.65\columnwidth]{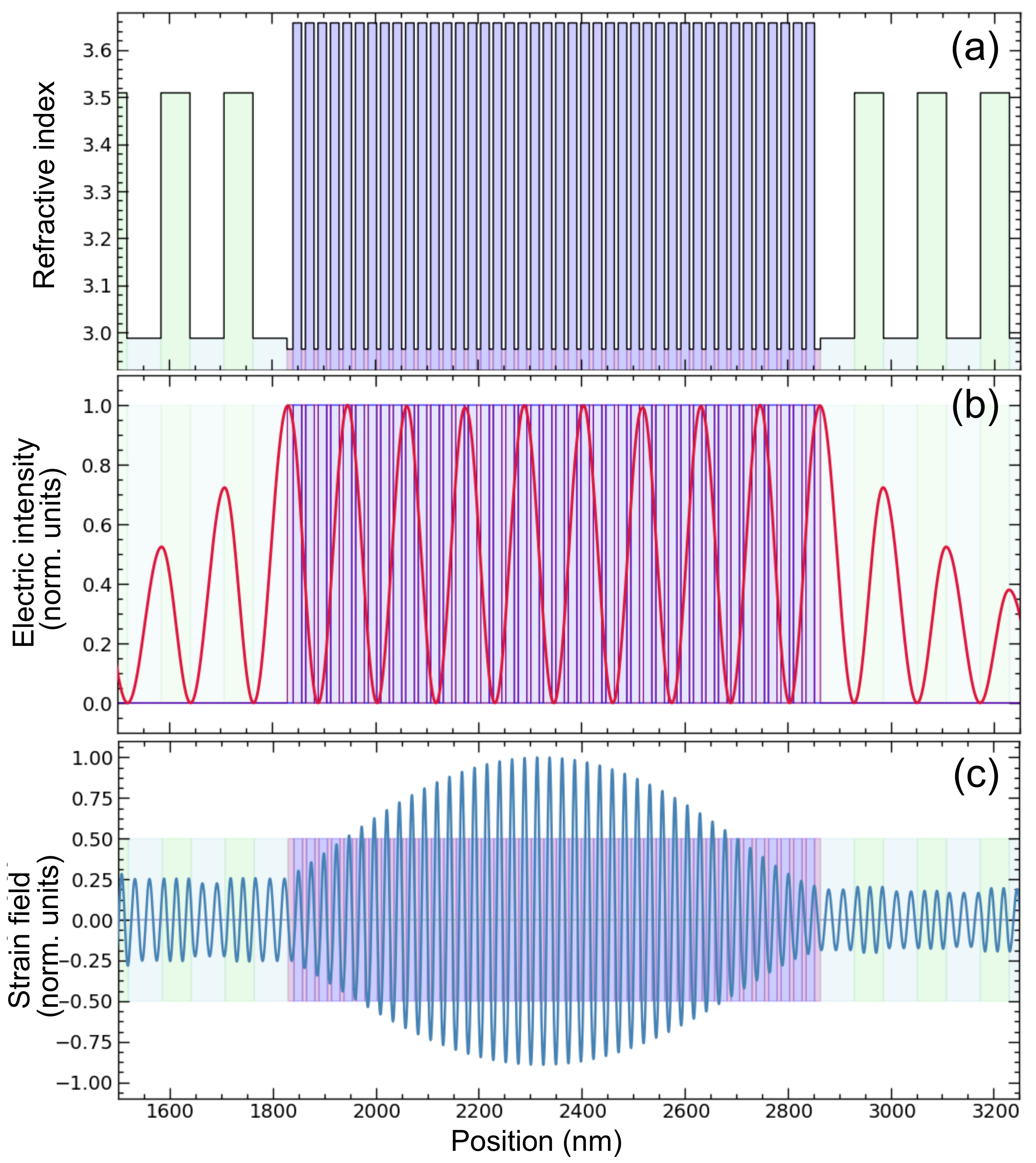}
%\vspace{-0.8 cm}
\caption{(a)  Refractive index profile around the spacer region, plotted along the growth direction of sample B. The  embedded quantum wells are shaded in purple. (b) Normalized intensity of the resonant electric field at 1.524 eV. (c) Normalized strain field of the $\sim 180$ GHz phonon mode. }
\label{FigS02}
\end{center}
\end{figure}

Sample B is a planar, $9\lambda/2$ semiconductor microcavity with a spacer constituted by a 41.5-period multiple quantum well (MQW). In Fig. \ref{FigS02}(a) the refractive index profile around this region is plotted. In Fig. \ref{FigS02}(b) the normalized optical intensity for the resonant wavelength is plotted in a solid red line.  The top (bottom) optical DBR is formed by 15 (19) periods of Al$_{0.1}$Ga$_{0.9}$As/Al$_{0.95}$Ga$_{0.05}$As $\lambda/4$ layers. The embedded MQW is identical to the bare one studied in Ref. \cite{SM_Jusserand} without optical confinement. It consists of 41 17.1-nm GaAs quantum wells separated by 42 AlAs barriers of 7.5 nm of thickness. The electronic states in each quantum well have a negligible overlap with the neighbors and thus lead to localized excitons only radiatively coupled. Relatively thick QWs were chosen to minimize the inhomogeneous broadening and dephasing. The DBRs were designed to match the photonic linewidth to the excitonic linewidth at low temperatures. This contributes to both achieving a higher Rabi gap and maintaining the lifetime of the lower polariton relatively constant across the whole detuning range \cite{SM_Rozas2014}. The exciton energy depends on temperature and in sample B it varies from $\sim 1.434\,$eV to $\sim 1.531\,$eV, as temperature changes shift from $\sim 293\,$K to $\sim 5\,$K. Thus, the exciton-polariton detuning, defined as difference between the non-interacting cavity energy and heavy-hole exciton energy ($E_c - E_{hh}$), can be set to any value ranging from $\sim -100\,$meV to $20\,$meV. Also, the resulting optical quality factor is  $Q_\textrm{opt}\sim 3\times 10^4$ in the purely-photonic condition. 

The MQWs function as also periodically modulated acoustic structure, so their acoustic properties are described in the folding scheme \cite{SM_Jusserand-LightScattSol5(1989)}. Acoustic waves with wavevector shifted by a multiple of $2\pi/d$ (where d is the thickness of a single QW period), and frequencies much larger than the standard Brillouin frequency thus becoming active for polariton scattering. The whole system (MQW+optomechanical microcavity) allows for the coupling between polaritons and 180 GHz zero group velocity (ZGV) phonon mode. The strain field related to the ZGV phonon mode is plotted in Fig. \ref{FigS02}(c) in solid blue lines. This acoustic zone-center mode has almost zero group-velocity because it is located at the lower edge of the first phononic bandgap of the MQW (see Sec. \ref{sec: Raman spectra in the cavity - phonon peaks} for details). This slow speed leads to an effective mechanical quality factor {estimated to be} $Q_m \sim 10^3$. {The latter is the nominal value obtained from simulations (see Sec. \ref{sec: Raman spectra in the cavity - phonon peaks}). The value obtained through measurements of the Brillouin phonon linewidth using the high-resolution mode of the spectrometer in its triple-additive configuration is $\sim$4$\times$10$^2$, and it is resolution limited.}

%%%%%%%%%%%%%%%%%%%%%%%%%%%%%%%%%%%%%%%%%%%%%%%%%%%%%%%%%%%%%%%%%%%%%%%%%%%

%%%%%%%%%%%%%%%%%%%%%%%%%%%%%%%%%%%%%%%%%%%%%%%%%%%%%%%%%%%%%%%%%%%%%%%%%%%
%\clearpage
%%%%%%%%%%%%%%%%%%%%%%%%%%%%%%%%%%%%%%%%%%%%%%%%%%%%%%%%%%%%%%%%%%%%%%%%%%%%%%%%%%%%%%%%%%%%%%%%%%%%%%%%%%%%%%%%%%%%%%%%%%%%%%%%%%%%%%%%%%%%%%%%%%%%%%
%%%%%%%%%%%%%%%%%%%%%%%%%%%%%%%%%%%%%%%%%%%%%%%%%%%%%%%%%%%%%%%%%%%%%%%%%%%%%%%%%%%%%%%%%%%%%%%%%%%%%%%%%%%%%%%%%%%%%%%%%%%%%%%%%%%%%%%%%%%%%%%%%%%%%%
\section{Experimental details}\label{secExperimentalDetails}
\begin{figure}[hhh!!!]
\begin{center}
\includegraphics*[keepaspectratio=true, clip=true, trim = 0mm 0mm 0mm 0mm, angle=0, width=0.45\columnwidth]{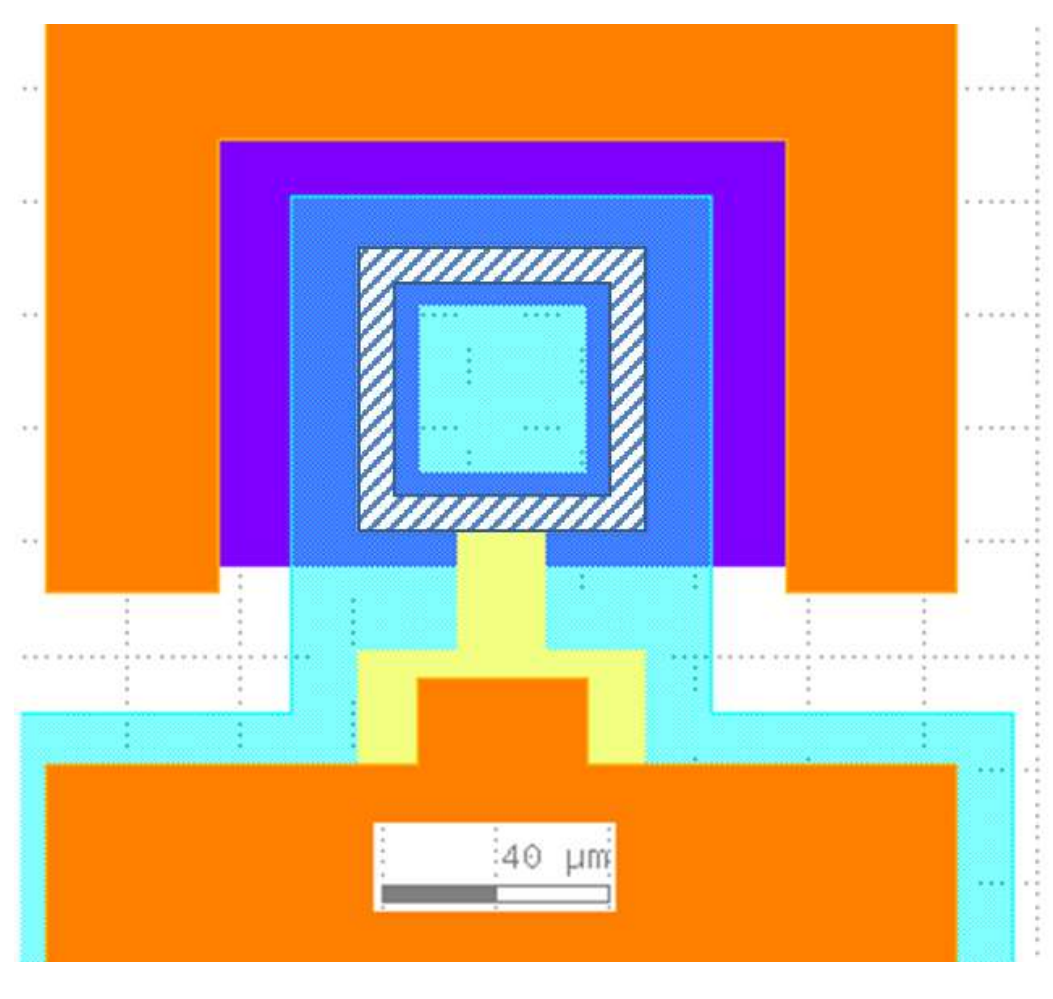}
%\vspace{-0.8 cm}
\caption{A schematic top view of a BAWR. The hatched area with the aperture is the active area that generates phonons. More details about BAWR fabrication and characterization can be found in Ref. \cite{SM_Machado}.}
\label{FigS03}
\end{center}
\end{figure}

The optomechanical coupling factor in sample A was determined by measuring the energy modulation amplitude of polariton emission, under driving by electrically generated bulk acoustic strain. The strain was injected into the spacer of the cavity using bulk acoustic wave resonators (BAWR) fabricated on the upper surface of the microcavity \cite{SM_Machado}. The active area of the transducer has a square-shaped aperture for optical access [see Fig. \ref{FigS03}]. Upon electrical excitation around $7\,$GHz, BAWRs generate longitudinal bulk acoustic waves of the same frequency that propagate nearly perpendicular to the cavity surface. Several BAWRs were fabricated at different positions on the tapered sample, which enabled to probe different exciton-photon detunings.

In a typical experiment, a non-resonant diode laser at $635\,$nm (and average optical power of $0.2\,$mW) was focused on a $10\,\mu$m$^2$ spot, in the center of the BAWR aperture. The laser pulses have a repetition rate of $30\,$MHz, and widths far exceeding the typical photon lifetimes in the microcavity. The PL is spectrally resolved by means of a single grating spectrometer with a resolution of $150\,\mu$eV and detected by a liquid-nitrogen-cooled CCD detector. Pulsing the laser in and out-of phase with the RF signal driving the BAWR, we verified the negligible influence of RF-induced heating on the polariton PL \cite{SM_Kuznetsov2021}.

%%%%%%%%%%%%%%%%%%%%%%%%%%%%%%%%%%%%%%%%%%%%%%%%%%%%%%%%%%%%%%%%%%%%%%%%%%%

\begin{figure}[hhh!!!]
\begin{center}
\includegraphics*[keepaspectratio=true, clip=true, trim = 0mm 0mm 0mm 0mm, angle=0, width=0.55\columnwidth]{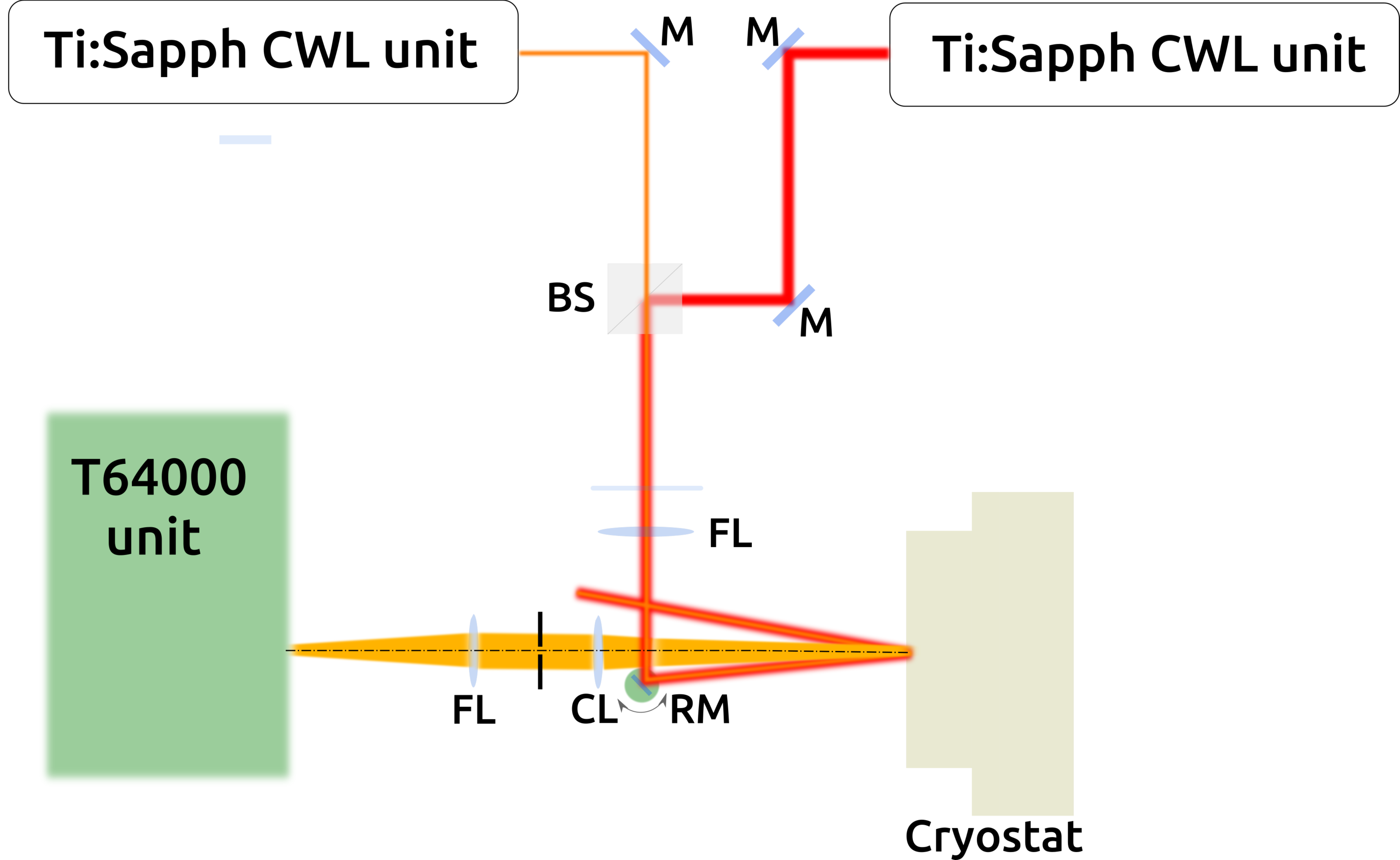}
%\vspace{-0.8 cm}
\caption{Experimental setup for simultaneous photoluminescence and Brillouin scattering spectroscopy performed in sample B (see text for more details).}
\label{FigS04}
\end{center}
\end{figure}

Sample B is studied by means of photoluminescence and Brillouin scattering spectroscopy using a custom-made macro configuration. This setup allows focusing two \textit{CW} titanium-sapphire lasers on perfectly overlapping spots {($\O \sim$20\,$\mu$m)}. This setup is presented in Fig. \ref{FigS04}. The first one (red in Fig. \ref{FigS04}) is used to excite the MQW photoluminescence when required to identify the spectral position of the polaritonic modes. This laser is tuned to the DBR's edge mode, above the excitonic energy to efficiently access the MQW with absorption only in the GaAs layers, and locally excite its photoluminescence. This laser is used at very low power (typically some $\mu$W), and it is turned off for Brillouin scattering experiments. It should be noted that, for the extraction of real polaritonic linewidths by photoluminescence spectroscopy, it is necessary to avoid large spot size, since it introduces a broadening due to the sample's thickness gradient.

On the other hand, Brillouin scattering spectroscopy is performed using the second single-mode stabilized titanium-sapphire laser (orange in Fig. \ref{FigS04}). The Stokes scattering process of the MQW's vibrational modes is studied in double optical resonance (DOR) \cite{SM_FainsteinPRL1995}, pumping and detecting through the lower cavity polariton branch (LP). It is performed with the laser excitation set in angle and collecting the emitted photons in the normal direction of the sample \cite{SM_FainsteinPRL1995}.   The scattering processes  through this branch is studied from a purely photonic condition to an excitonic character condition. The power of the second laser is varied from $~50\,$mW in the photonic condition to some $\mu$W in the excitonic condition. Its energy is varied from $\sim$1.39\,eV to $\sim$1.53\,eV. In all cases, excitation is well below the gaps of the DBRs and MQW's barriers, and thus its contribution to additional heating is negligible. Also, although zero-wavevector phonons, like the studied slow ZGV MQW mode, would require in principle a forward-scattering geometry to be coupled by inelastic light scattering, they are readily accessed when exciting through an optical cavity mode in a back-scattering configuration, as schematized in Fig \ref{FigS04}. 
%%%%%%%%%%%%%%%%%%%%%%%%%%%%%%%%%%%%%%%%%%%%%%%%%%%%%%%%%%%%%%%%%%%%%%%%%%%
%%%%%%%%%%%%%%%%%%%%%%%%%%%%%%%%%%%%%%%%%%%%%%%%%%%%%%%%%%%%%%%%%%%%%%%%%%%
{For all experiments in sample B a high-resolution spectrometer Jobin\,Yvon--Horiba T64000, equipped with a liquid nitrogen CCD, was used to acquire PL and BS spectra with a resolution of $\sim 3\,$GHz. The sample is located in a cryostat to vary the temperature from 5 to 293K, with a PID control system. 	}\\

%%%%%%%%%%%%%%%%%%%%%%%%%%%%%%%%%%%%%%%%%%%%%%%%%%%%%%%%%%%%%%%%%%%%%%%%%%%
%\clearpage
%%%%%%%%%%%%%%%%%%%%%%%%%%%%%%%%%%%%%%%%%%%%%%%%%%%%%%%%%%%%%%%%%%%%%%%%%%%%%%%%%%%%%%%%%%%%%%%%%%%%%%%%%%%%%%%%%%%%%%%%%%%%%%%%%%%%%%%%%%%%%%%%%%%%%%
%%%%%%%%%%%%%%%%%%%%%%%%%%%%%%%%%%%%%%%%%%%%%%%%%%%%%%%%%%%%%%%%%%%%%%%%%%%%%%%%%%%%%%%%%%%%%%%%%%%%%%%%%%%%%%%%%%%%%%%%%%%%%%%%%%%%%%%%%%%%%%%%%%%%%%
 
\section{Theoretical aspects}\label{secTheory}
In this section, the theoretical models used to evaluate the optomechanical coupling and Brillouin intensity with varying resonant conditions (detuning) are presented. 

\subsection{Spectral densities of polariton modes}\label{Spectral_densities}
Cavity polaritons correspond to coupled states involving cavity photons and QW-confined excitons with the same in-plane wave vector. Thus, the system is readily modeled as three oscillators, and its non-interacting hamiltonian in terms of $\hbar$ reads
\begin{eqnarray}\label{eqn: Hamiltonian0}
\hat{\mathcal{H}}_{\textrm{0}}= \omega_c \hat{c}^{\dagger} \hat{c}+ \omega_{hh} \hat{a}^{\dagger}_{hh} \hat{a}_{hh} + \omega_{lh} \hat{a}^{\dagger}_{lh} \hat{a}_{lh}, 
\end{eqnarray}
%\mbox{$\frac{1}{2}$}
where $\omega_c$, $\omega_{hh}$, and $\omega_{lh}$ correspond to the complex energies for cavity photons, heavy-hole-like and light-hole-like excitons.  The decay rate for them is  contemplated including an imaginary part \cite{SM_Axel} as follows: $\omega_c \to \omega_c-i\Gamma_c/2$, $\omega_{hh}\to \omega_{hh}-i\Gamma_{hh}/2$, and $\omega_{lh}\to \omega_{lh}-i\Gamma_{lh}/2$. 

The interaction between confined photons and excitons is expressed by
\begin{eqnarray}\label{eqn: Hamiltonian1}
\hat{\mathcal{H}}_{\textrm{int}}= \frac{\Omega_{hh}}{2} (\hat{c}^{\dagger} \hat{a}_{hh} + \hat{a}^{\dagger}_{hh} \hat{c}) + \frac{\Omega_{lh}}{2} (\hat{c}^{\dagger} \hat{a}_{lh} + \hat{a}^{\dagger}_{lh} \hat{c}).
\end{eqnarray}

\begin{figure}[hhh!!!]
\begin{center}
\includegraphics*[keepaspectratio=true, clip=true, trim = 0mm 0mm 0mm 0mm, angle=0, width=0.75\columnwidth]{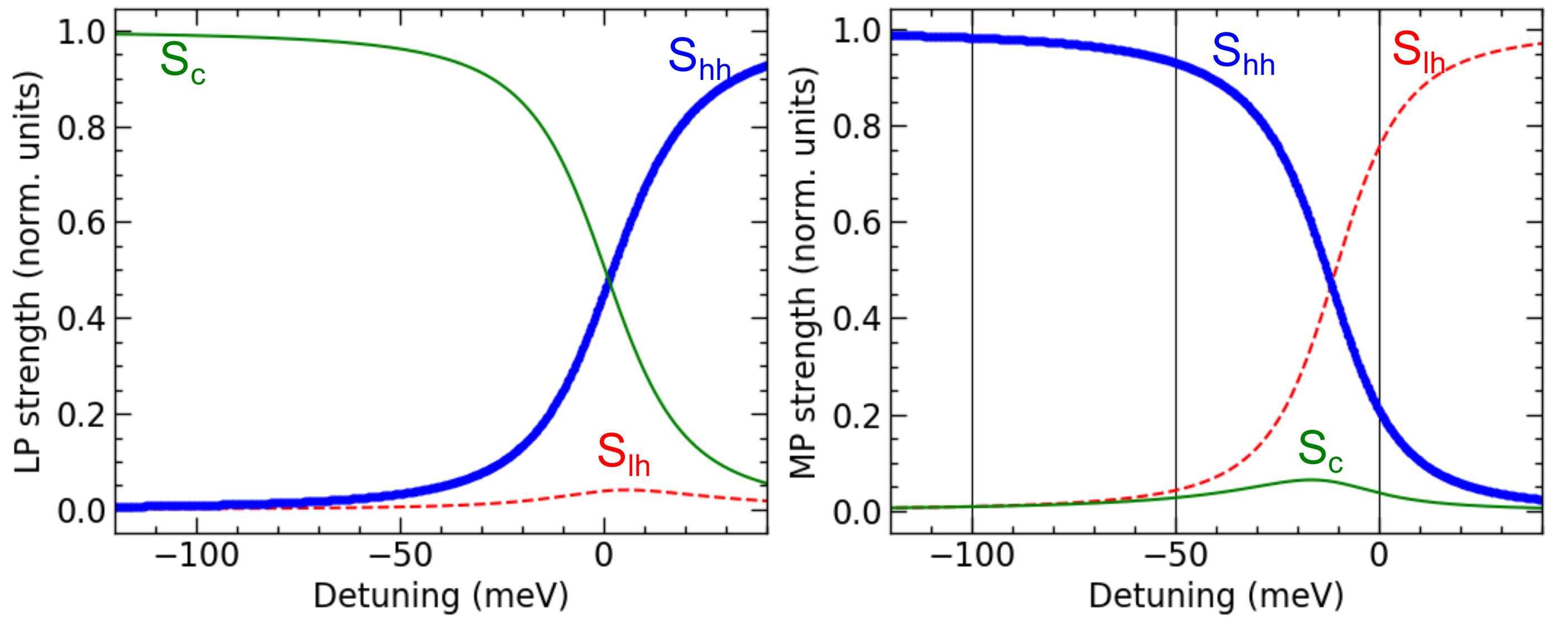}
%\vspace{-0.8 cm}
\caption{Hopfield coefficients for lower (left pannel) and medium (right pannel) polaritons calculated for sample A (used in Fig. 1(d) in main text). Heavy-hole and light-hole excitonic strength of these branches, $S_{hh}$ and $S_{lh}$, are plotted with solid blue and dashed red lines, respectively. Photonic strength $S_c$ for LP and MP is presented with solid green lines.}
\label{FigHOP}
\end{center}
\end{figure}

This problem is solved exactly. We proceed to calculate the retarded Green functions associated with the photons and excitons number operators. In order to do that, the spectral equation of motion for any operators $\hat{p}$ and $\hat{q}$ is considered
\begin{eqnarray}\label{eqn: EqOfMotionRGF}
\omega \langle \langle \hat{p}, \hat{q} \rangle \rangle_\omega = \langle [\hat{p}, \hat{q}] \rangle + 
\langle \langle [\hat{p}, \hat{\mathcal{H}}], \hat{q} \rangle \rangle_\omega,
\end{eqnarray}
where $\langle \langle , \rangle \rangle_\omega$ denotes the spectral retarded Green function for the considered operators, and $[,]$ denotes the bosonic commutator operator. 
Thus, the retarded Green function for photons reads
\begin{eqnarray}\label{eqn: cc}
\langle \langle \hat{c}, \hat{c}^{\dagger} \rangle \rangle_\omega = \frac{1}{\omega
- \omega_c + i\frac{\Gamma_c}{2} - \frac{\big(\frac{\Omega_{hh}}{2}\big)^2}{\omega - \omega_{hh}  + i\frac{\Gamma_{hh}}{2}} - \frac{\big(\frac{\Omega_{lh}}{2}\big)^2}{\omega - \omega_{lh}+ i\frac{\Gamma_{lh}}{2}}},
\end{eqnarray}
while for excitons corresponds to
\begin{eqnarray}\label{eqn: x1x1}
\langle \langle \hat{a}_{hh}, \hat{a}^{\dagger}_{hh} \rangle \rangle_\omega = \frac{1}{\omega
- \omega_{hh} + i\frac{\Gamma_{hh}}{2} - \frac{\big(\frac{\Omega_{hh}}{2}\big)^2}{\omega - \omega_{c} + i\frac{\Gamma_c}{2}-\frac{\big(\frac{\Omega_{lh}}{2}\big)^2}{\omega - \omega_{lh}+ i\frac{\Gamma_{lh}}{2}}}} 
\end{eqnarray}
for heavy-hole excitons and 
\begin{eqnarray}\label{eqn: x2x2}
\langle \langle \hat{a}_{lh}, \hat{a}^{\dagger}_{lh} \rangle \rangle_\omega = \frac{1}{\omega
- \omega_{lh} + i\frac{\Gamma_{lh}}{2} - \frac{\big(\frac{\Omega_{lh}}{2}\big)^2}{\omega - \omega_{c} + i\frac{\Gamma_c}{2}-\frac{\big(\frac{\Omega_{hh}}{2}\big)^2}{\omega - \omega_{hh}+ i\frac{\Gamma_{hh}}{2}}}}
\end{eqnarray}
for light-hole excitons. The spectral density for photons [$A_c(\omega)$] and excitons [$A_{hh}(\omega)$ and $A_{lh}(\omega)$] are determined by %taking the imaginary part of their retarded Green functions as follows,%, and the number of photons and excitons can be obtained by integrating them. 
\begin{eqnarray}\label{eqn: Ac}
A_c(\omega)=-\frac{1}{\pi}Im\langle \langle \hat{c}, \hat{c}^{\dagger} \rangle \rangle_\omega,
\end{eqnarray}
\begin{eqnarray}\label{eqn: Ax1}
A_{hh}(\omega)=-\frac{1}{\pi}Im\langle \langle \hat{a}_{hh}, \hat{a}^{\dagger}_{hh} \rangle \rangle_\omega, 
\end{eqnarray}
and
\begin{eqnarray}\label{eqn: Ax2}
A_{lh}(\omega)=-\frac{1}{\pi}Im\langle \langle \hat{a}_{lh}, \hat{a}^{\dagger}_{lh} \rangle \rangle_\omega. 
\end{eqnarray}
By integrating these spectral densities over each branch, the corresponding photonic ($S_{c}$) and excitonic strengths  ($S_{hh}$, and $S_{lh}$) are determined. These densities reflect which component is present at energy $\omega$ for each polariton mode. The excitonic and photonic strengths for the LP and MP modes of sample A, are presented on the left and right panels of  Fig. \ref{FigHOP}, respectively.

%%%%%%%%%%%%%%%%%%%%%%%%%%%%%%%%%%%%%%%%%%%%%%%%%%%%%%%%%%%%%%%%%%%%%%%%%%%
%%%%%%%%%%%%%%%%%%%%%%%%%%%%%%%%%%%%%%%%%%%%%%%%%%%%%%%%%%%%%%%%%%%%%%%%%%%

\subsection{Photoluminescence experiments}\label{secTheoryPL}
Cavity polaritons are observed through photoluminesce experiments. Optically pumping the system with energy higher than the bandgap energy of the active materials, at very low power (typically $\sim \mu$W--mW), polaritons are detected in the outside world through their photonic character. Therefore, the photons detected in the photoluminescence experiment are considered to be proportional to the photonic density of states of polaritons.  
%%%%%%%%%%%%%%%%%%%%
In the negative detuning limit the lower polariton is purely photonic and the decay rates are less dependent on temperature (shown in Fig. 3b in the main text). Thus, the intrinsic photonic decay rate is fixed to $0.40\,$meV. The excitonic decay rate is fitted for different temperatures (shown in Fig. 3c in the main text).

\subsection{Brillouin Intensity}\label{secTheoryBL}
The outgoing Brillouin scattering intensity mediated by lower polaritons is calculated in the same spirit of Ref. \cite{SM_Rozas2014} within the ``factorization model'' of polariton mediated scattering described as follows: i) a photon of energy $\omega_i$ impinges on the cavity and is converted onto a polariton $P_i$ with probability $T_i$. ii) Polariton $P_i$ interacts with phonons through two different channels. The excitonic component provides the more significant coupling through deformation potential. Also, the photonic component supports the coupling with phonons through radiation-pressure and non-resonant photoelastic mechanisms. iii) The scattered polariton $P_s$ leaves the system with energy $\omega_s$ with a probability $T_s$.
The Brillouin scattering intensity is then, 
\begin{eqnarray}\label{eqn: Raman}
I_B \propto T_i W_{i \to s} T_s,
\end{eqnarray}
where $T_i$ and $T_s$ describe the incoming and outgoing channels, respectively, which are proportional to the photonic character of the involved polaritons. On the other hand, $W_{i \to s}$ is the scattering probability from the initial state (i) to the final polariton state (s) per unit of time,
\begin{eqnarray}\label{eqn: Raman2}
W_{i \to s} = \frac{2 \pi}{\hbar}\| \langle i | H' | s \rangle \|^2 \rho(\omega).
\end{eqnarray}
Note that the effects of the lifetime of the outgoing polariton $P_s$ are included in the final density of states $\rho(\omega)$.
The polariton-phonon interaction $H'$ is proportional to the excitonic part of the scattered polariton through resonant deformation potential mediation, scaled by the excitonic strength $S^{LP}_X$, and to the photonic part of the polariton through radiation-pressure mediation scaled by the photonic strength $S^{LP}_c$. Thus, we express the Brillouin intensity $I_B$ as
\begin{eqnarray}\label{eqn: Raman3}
I_B \propto  S_c^{LP_i} \Gamma_{LP_i}^{-1} (S_{hh}^{LP}  + S_{lh}^{LP} +a S_c^{LP} )^2 S_c^{LP_s} \Gamma_{LP_s}^{-1},  
\end{eqnarray}
where $\Gamma_{LP_i}$, and $\Gamma_{LP_s}$ are the initial and final lower polariton linewidths, respectively. Here it is assumed that the phonon frequency is small, so that the Hopfield coefficients corresponding to incoming and scattered polaritons are the same. 
The factor $a=0.004$ takes into account the ratio between the DP and RP coupling, which for sample B was fitted to be $a \sim 0.004$. The temperature effects in the intensity are included through the impact on the polariton linewidth (see Fig. 3(b) on main text) and Rabi splittings.  
The double-optical resonant condition $\omega_i - \omega_s' = \Omega_M$ is set, $\Omega_M$ being the phonon energy \cite{SM_FainsteinPRL1995}.

\subsection{Optomechanical coupling}\label{secOMCF}
The methodology to determine the optomechanical coupling rates in 1D semiconductor microcavities is now introduced.
First, a generic multilayered semiconductor nanostructure is considered, which is characterized by its axis-symmetrical growth direction $\hat{z}$. Any optical or mechanical magnitude in the system is written in terms of multilayered crystal. Each layer in the structure is represented by an integer $j$. Furthermore, the distribution functions for basic optical and mechanical properties are presented in Fig. \ref{FigS01}(a) and Fig. \ref{FigS02}(a), where the embedded MQW inside the microcavity is schematized, and the real part of refractive index profile is plotted  (black solid line). Different materials in the structure are shaded with distinct colors in Figs. \ref{FigS01} and \ref{FigS02}. Green and purple layers represent GaAs wells in sample A and B, respectively.

Photons and phonons are treated as propagating fields, which are described by wave equations. Considering an homogenous and isotropic non-magnetic medium, the electric field propagation along the $\hat{z}$ direction is given by the expression  
\begin{equation}
\label{Electric}
\frac{\partial^2 \vec{E}}{\partial z^2} - \frac{1}{v_L^2}\frac{\partial^2 \vec{E}}{\partial t^2}=0, 
\end{equation}
where $\vec{E}$ corresponds to the in-plane component of the electric field, and $v_L = c/n $ is the phase velocity in each layer, and $c$ the speed of light in vacuum \cite{SM_Jackson}.

For vibrations in the same layer,  the displacement field is described by the vector $\vec{u}$, which is governed by a similar wave equation,
\begin{equation}
\label{Mechanical}
\frac{\partial^2 \vec{u}}{\partial z^2} - \frac{1}{v_s^2}\frac{\partial^2 \vec{u}}{\partial t^2}=0,
\end{equation}
where $v_s=\sqrt{\mathcal{C}/\rho}$ is the mechanical wave velocity along $\hat{z}$, $\mathcal{C}$ the elasticity modulus, and $\rho$ the density of the material. In expressions (\ref{Electric}) and (\ref{Mechanical}) we have assumed that the propagating waves are independent of other directions. Let us consider now a stack of different non-magnetic homogenous media. When an interface between two different material layers is considered at a certain position $z_0$, the solution for the wave propagating field has also to match appropriate boundary conditions.
In a more general way, both cases studied are represented by a wave-equation of the form
\begin{equation}
\label{general}
\frac{\partial^2 X}{\partial z^2} - \frac{1}{v_s^2}\frac{\partial^2 X}{\partial t^2}=0,
\end{equation}
whose solution is written as
\begin{equation}
\label{solucionX}
X_j(z_j,t) = \big( a_j e^{i k_j z_j} + b_j e^{-i k_j z_j} \big) e^{i \omega t},
\end{equation}
and the boundary conditions at the interface between two consecutive layers are expressed as follows
\begin{equation}
\label{transfer-matrix}
\left( \begin{array}{c}
a_{j+1}\\
b_{j+1}
\end{array} \right) = \frac{1}{2}
\left( \begin{array}{cc}
(1 + Z_{j}) e^{i k_j d_j} & (1 - Z_{j}) e^{-i k_j d_j}  \\
(1 - Z_{j}) e^{i k_j d_j} & (1 + Z_{j}) e^{-i k_j d_j}
\end{array} \right)
\left( \begin{array}{c}
a_{j}\\
b_{j}
\end{array} \right),
\end{equation}
where the index $j$ labels the $j$-th layer in the structure, $k_j=\omega/v_j$, and $Z_j=n_j/n_{j+1}$ for the optical field and  $Z_j=\rho_j v_j /\rho_{j+1} v_{j+1}$ for the acoustic field. The coefficients $a_j$ and $b_j$ are defined by the boundary conditions. For photons they correspond to $a_0 = 1$, $b_N = 0$. For the case of phonons, stress-free boundary conditions are imposed at surface and backside of the substrate, and then the field is divided by its maximum $|u_0|$. 

The dielectric function of each semiconductor material is expressed, for simplicity, to be constant in energy. This is valid at energies far from resonances. This approximation is valid for all but GaAs layers. In the latter, the dielectric function will be considered as the background, plus Lorentz oscillators taking into account the excitonic effects in the system \cite{SM_Chen1995}. Thus, for the optically active layers (GaAs), the dielectric function reads
\begin{equation}
\label{EQ01}
\varepsilon(\omega) = \varepsilon_\infty + \sum_l \frac{4\pi \beta \omega_l^2 }{\omega_l^2 - \omega^2 - i \Gamma_l \omega},
\end{equation}
where $\varepsilon_\infty$ is the background permittivity, $\omega_l$ is the l-th excitonic 
energy and $\Gamma_l$ corresponds to the l-th excitonic linewidth. Typical values for all these parameters are extracted from PL spectra. Note that all of the parameters depend on temperature. The factor $4\pi \beta \omega_l^2$ corresponds to the strength of these oscillators and is equal to half of the Rabi Splitting $\Omega_{Rj}$.
%%%%%%%%%%%%%%%%%%%%%%%%%%%%%%%%%%%%%%%%%%%%%
\begin{figure}[!hht]
 \begin{center}
\includegraphics*[keepaspectratio=true, clip=true, trim = 0mm 0mm 0mm 0mm, angle=0, width=0.6\columnwidth]{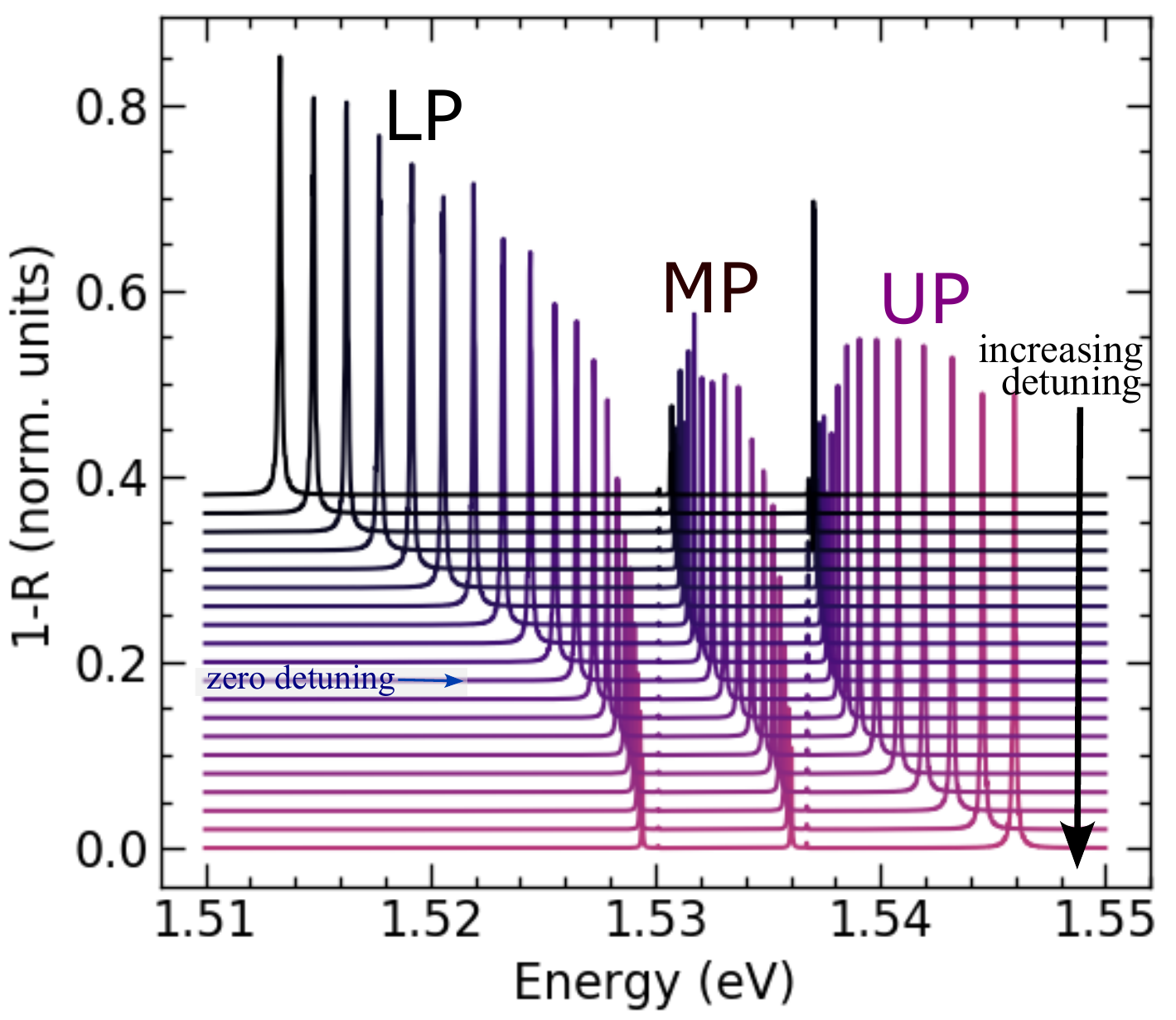}
\end{center}
%\vspace{-0.8 cm}
%    \hspace{-1.6 cm}
\caption{Modeled optical response for the exciton-polariton system at different detunings (the spectra are vertically shifted for clarity). Mainly, three peaks are observed in a stop-band flat spectral region, which corresponds to the lower (LP), medium (MP), and upper (UP) polariton lines, respectively.
}
\label{Polariton}
\end{figure}
%%%%%%%%%%%%%%%%%%%%%%%%%%%%%%%%%%%%%%%%%%%%%
The optical response of the described system, monitored through the reflectance $R = |b_0|^2$,  is plotted in Fig. \ref{Polariton} using the low-temperature parameters mentioned above. The spectra in Fig. \ref{Polariton} are plotted for different cavity thicknesses, leading from positive to negative detunings, vertically shifted for clarity. Although transmittance $T$ is small because of the residual absorption of the thick substrate, is taken into account to determine absorption $A=1-R-T$.  The range evaluated corresponds to the stop-band high reflectance region, in which the characteristic avoided crossing between the photonic and excitonic states appears. The lower energy branch (LP) will be studied in the presence of a frozen phonon field to evaluate its optomechanical coupling rate. This lower polariton state is coupled with phonons with its photonic character in the region of transparency via radiation-pressure and non-resonant photoelastic mechanism. Also, LP allows to exploit a resonant coupling with phonons via deformation-potential interaction. We calculate the optical response of the system in the presence of confined acoustic phonons. The optomechanical linear coupling is tested including radiation-pressure and resonant photoelastic mechanism. For radiation pressure, all the interfaces of the structure are perturbed by the displacement field $u(z)$, which is modulated for different amplitudes. The deformation-potential interaction is accounted including excitonic energy shifts proportional to the strain field $s=\partial u/\partial z$ as $\tilde{\omega}_{xj} = \omega_{xj} - \Xi_{xj} \times s$. Following the shift of the LP energy as a function of the phonon amplitude, we determine the linear optomechanical coupling coefficient.
The only parameters used are  $\Xi_{hh}=-10.5\,$eV and $\Xi_{lh}=-6.5\,$eV, and typical values for refractive indices, densities, and sound velocities for the constituent materials presented in Table \ref{tab1}.
\begin{table}
\begin{ruledtabular}
\begin{tabular}{c c c c}
Material & refractive index $n$ & speed of sound $v$ (m/s) & density $\rho$ (kg/m$^3$)\\
\hline
GaAs & 3.657 & 4727 & 5317 \\
Al$_{0.10}$Ga$_{0.90}$As & 3.509 & 4800 & 5161 \\
Al$_{0.15}$Ga$_{0.85}$As & 3.446 & 4844 & 5074 \\
Al$_{0.30}$Ga$_{0.70}$As & 3.296 & 4999 & 4792 \\
Al$_{0.70}$Ga$_{0.30}$As & 3.082 & 5449 & 4056 \\
Al$_{0.90}$Ga$_{0.10}$As & 3.005 & 5610 & 3816 \\
Al$_{0.95}$Ga$_{0.05}$As & 2.987 & 5637 & 3781 \\
AlAs & 2.964 & 5656 & 3757 \\
\hline
\end{tabular}
\end{ruledtabular}
\caption{Optical and mechanical properties for the materials used.}
\label{tab1}
\end{table} 

\subsection{Estimation of the electrically generated bulk acoustic wave strain}\label{BAWstrain}
 
To estimate the strain induced by the bulk acoustic waves (BAWs) on the embedded quantum wells (QWs), first the electrically coupled radio-frequency power is calculated as $P_{m}=r_{rf}t_{cable}P_{rf}$, where $P_{rf}$ is the applied electric power, $r_{rf}$ the coupling efficiency experimentally extracted from electrical reflectivity curves, (i.e., from the $s_{11}$ radio-frequency scattering parameter) and $t_{cable}$ are the cable losses ($\sim 6$~dB). The acoustic energy stored in the driven acoustic cavity mode is then evaluated as $E_m=P_m \tau_m$, with $\tau_m \sim 300$~ns the mode life-time experimentally derived from acoustic echo measurements.  The acoustic cavity mode has been observed to be characterized by a fine frequency comb~\cite{SM_Kuznetsov2021} implying that several reflections occur on the back and front surfaces of the wafer, which build up to define a Fabry-Perot-like standing acoustic wave. This also implies that a significant portion of the injected $rf$ power is located in the substrate, and not necessarily within the acoustic resonator.  Proper account of these effects is taken into considered by a phonon-propagation method, which includes acoustic reflections on the wafer boundaries.  Through this procedure, and assuming that the acoustic cavity mode stored energy is distributed uniformly over the $\sim 50 \mu$m diameter of the BAW resonator, we calculate that for $P_{rf}=0.1 W^{1/2}$ the strain at the QWs is $u_{zz} = 1.6 \times 10^{-3}$.

The fact that the BAW resonator has a ring-shape to allow for optical experiments implies a departure from the above calculation, which assumed that the BAW is uniformly distributed over the full area of the device. To account for these features, we have used diffraction theory to estimate the extension of the acoustic beam generated by a ring of $5 \mu$m width and $40 \mu$m central hole diameter, after propagating to the back side of the substrate and reflecting back to the surface, which sums to a total distance $z_{BAW}=700 \mu$m. We estimate a diffracted BAW beam-width $x_{BAW} \sim 30 \mu$m, which indeed is comparable to the ring diameter, thus implying that the assumption of an energy density approximately constant within the ring is sound. 
These results are in agreement with the results in Ref.~\onlinecite{Kuznetsov2021}, which show that the amplitude of the strain field within the ring-shaped transducer does not vary significantly. 
The experimental optomechanical couplings shown in Fig.~1(c) of the main text are thus obtained taking $u_{zz} = 1.6 \times 10^{-3}$ at  $P_{rf}=0.1 W^{1/2}$. As a check of consistency we mention that using the heavy-hole exciton deformation potential we expect for such a strain an energy modulation for a pure excitonic level $\Delta E_{cal}\sim 15$~meV, while the measured value for the middle polariton in the extreme negative detuning (where this mode has almost a pure heavy-hole excitonic character) is $\Delta E_{exp} \sim 13$~meV. Notwithstanding this excellent agreement and the one found on comparing the experiments in Fig.~1(c) with the theory of Fig.~1(d) of the main text,  in view of the several assumptions made we believe that a systematic error of a factor of 2 should be allowed for on using our results.

\section{Phonon modes and Brillouin spectra}\label{sec: Raman spectra in the cavity - phonon peaks}
\begin{figure}[h]
\begin{center}
\includegraphics*[keepaspectratio=true, clip=true, trim = 0mm 0mm 0mm 0mm, angle=0, width=0.85\columnwidth]{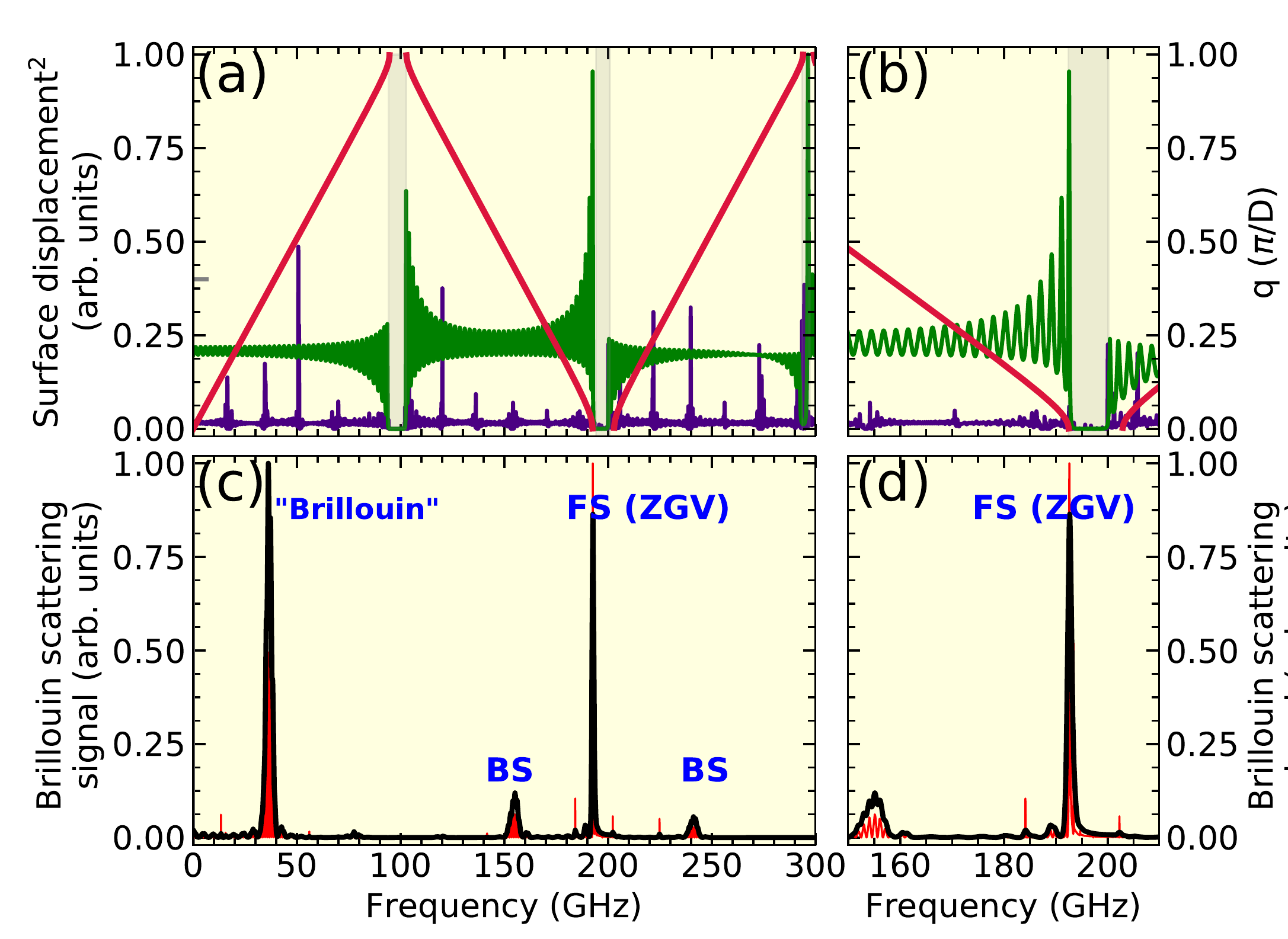}
	\caption{Acoustic surface displacement and Brillouin scattering spectrum. The top panels display the calculated squared acoustic surface displacements for the bare MQW (green curve) and for the complete optical microcavity structure with the embedded MQW (violet). The red curve corresponds to the tilted infinite MQW's dispersion. Bottom panels: The red curve corresponds to the calculated Brillouin spectrum and the black curve indicates its convolution with the experimental resolution. Labels FS and BS indicate the acoustic modes that have $q\sim 0$ and $q\sim 2 k_L$, respectively.}\label{fig:Spectra}
\end{center}
\end{figure}

The vibrational modes of sample B deserve a separate and more elaborate introduction. They are rather complicated, rising from the fact that the structure consists of a MQW embedded in an optical microcavity that simultaneously acts as an acoustic cavity.
The vibrational modes are obtained using a 1D continuum elastic model based on a transfer-matrix formalism, as explained in Sec. \ref{secOMCF}.
One way to understand the phononic structure of the system is to analyze the surface displacement as a function of the phonon energy. The green line in Fig.\,\ref{fig:Spectra} (top panels) presents the surface displacement of a bare finite MQW, i.e. the embedded MQW of 41.5 periods \textit{without} the optical microcavity. Superimposed, the red line shows the tilted dispersion relation of an equivalent infinite MQW showing the folding of the acoustic branch, and the opening of the acoustic mini-gaps \cite{SM_Jusserand-LightScattSol5(1989)}. These phonon bandgaps are indicated by the gray-shaded areas. It is clear that for the finite MQW (green curve) the surface displacement at the mini-gap is reduced to zero. The top-right panel corresponds to a zoom around the energy of interest ($\sim$5.7\,cm$^{-1}$). The violet curve shows the calculated surface displacement for the complete structure, i.e. the MQW embedded within the optical cavity. The surface displacement behaves equivalently to the green curve (bare MQW) at the gray mini-gap regions, which means that this contribution is inherited from the MQW. Additional features (e.g. peaks) also arise, resulting from the acoustic contribution of the optical cavity spacer, as well as the contribution from the optical DBRs \cite{SM_FainsteinPRL2013}.

\begin{figure}[h]
\begin{center}
\includegraphics*[keepaspectratio=true, clip=true, trim = 0mm 0mm 0mm 0mm, angle=0, width=0.95\columnwidth]{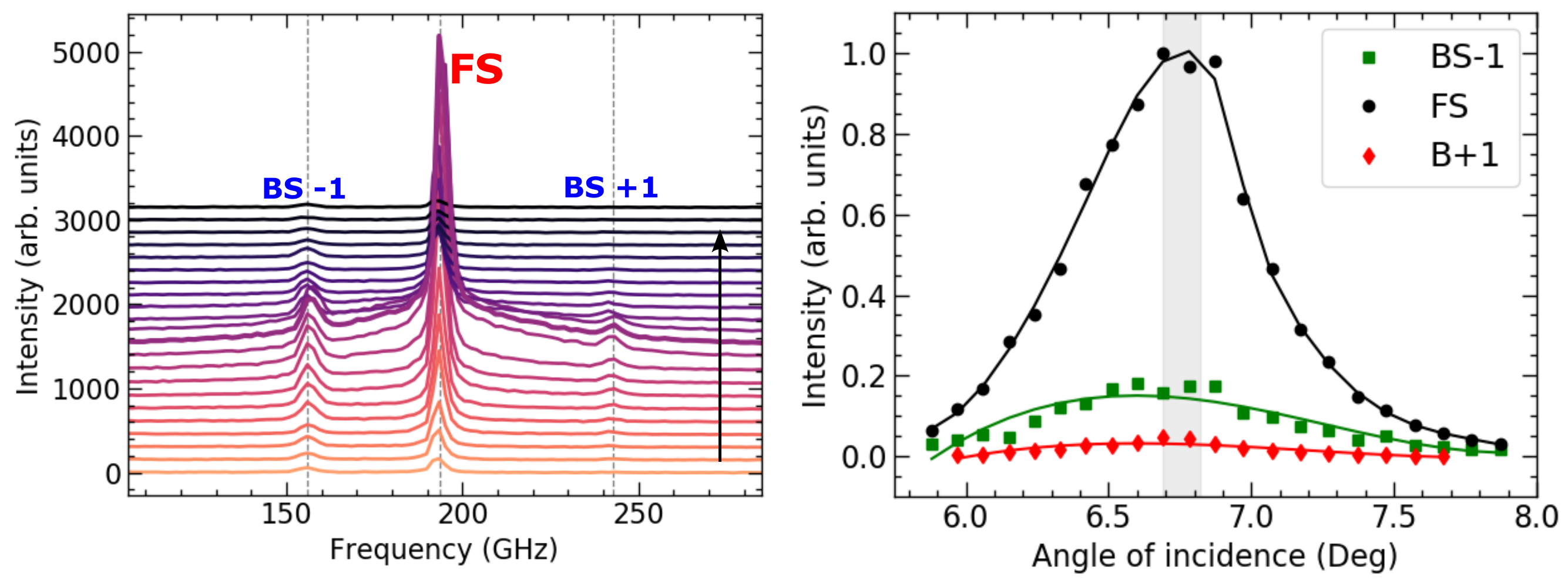}
	\caption{(left) Brillouin scattering spectra for different backscattering angle configurations at $78\,$K. (right) Angular dependence of  the intensity for the couple modes. The shaded region indicates the DOR condition.}\label{fig:DOR}
\end{center}
\end{figure}

The calculated Brillouin scattering spectrum is shown in Fig.\ref{fig:Spectra}$\,$(c) and (d). The one-dimensional simulation uses classical electrodynamics and a continuum elastic theory, and describes all the involved processes: incident photon; the creation/annihilation of the acoustic phonons; and the scattered electric field. The main ingredients are the overlap integral of the incoming electric field, the involved acoustic strain field $s(z)$, and the outgoing (scattered) electric field (see Fig. \ref{FigS02}). The red lines in Fig.\,\ref{fig:Spectra} (bottom panels) are the calculated Brillouin scattering intensities, and the black lines correspond to the convoluted signal with the experimental resolution ($\sim 3\,$GHz) of the spectrometer. From this result (see the detail in the right bottom panel) it is quite clear that the measured peaks are resolution limited (see Fig. \ref{fig:DOR}). 

The overall spectrum reveals the Brillouin peak at the lowest relative energy, together with the well-known folded acoustic (FA) triplets near the MQW's zone-center \cite{SM_Jusserand-LightScattSol5(1989)}. The Brillouin peak is not studied in our experiments. The peaks indicated as BS (back-scattering) have an acoustic wavevector $q\simeq 2 k_L$ ($k_L$ is the wavevector of the incident light), and the intense peak at $\sim 180\,$GHz (indicated as FS) is the slow Brillouin zone-center mode with $q\simeq 0$ (see e.g. Refs.\,\onlinecite{SM_Jusserand-LightScattSol5(1989)} for details). 
Brillouin triplets are observed experimentally as a function of the backscattering angle configuration. In Fig. \ref{fig:DOR} we present these triplets for different angles, recorded at $78\,$K in the purely photonic detuning. The spectra are vertically shifted for clarity (see the black arrow on the left pannel in Fig. \ref{fig:DOR}). The first central peak is the monitored ZGV mode intensity, also known as FS (forward scattering). On the right panel, the intensity for the triplet vs angular detuning. The double optical resonance condition ($\omega-\omega' = \Omega_M$) is defined by the angle of maximum intensity.

%\clearpage
%%%%%%%%%%%%%%%%%%%%%%%%%%%%%%%%%%%%%%%%%%%%%%%%%%%%%%%%%%%%%%%%%%%%%%%%%%%%%%%%%%%%%%%%%%%%%%%%%%%%%%%%%%%%%%%%%%%%%%%%%%%%%%%%%%%%%%%%%%%%%%%%%%%%%%
%%%%%%%%%%%%%%%%%%%%%%%%%%%%%%%%%%%%%%%%%%%%%%%%%%%%%%%%%%%%%%%%%%%%%%%%%%%%%%%%%%%%%%%%%%%%%%%%%%%%%%%%%%%%%%%%%%%%%%%%%%%%%%%%%%%%%%%%%%%%%%%%%%%%%%
\section{Optomechanical coupling factor in microcavity pillars}\label{g0}
\begin{figure}[h]
\begin{center}
\includegraphics*[keepaspectratio=true, clip=true, trim = 0mm 0mm 0mm 0mm, angle=0, width=0.45\columnwidth]{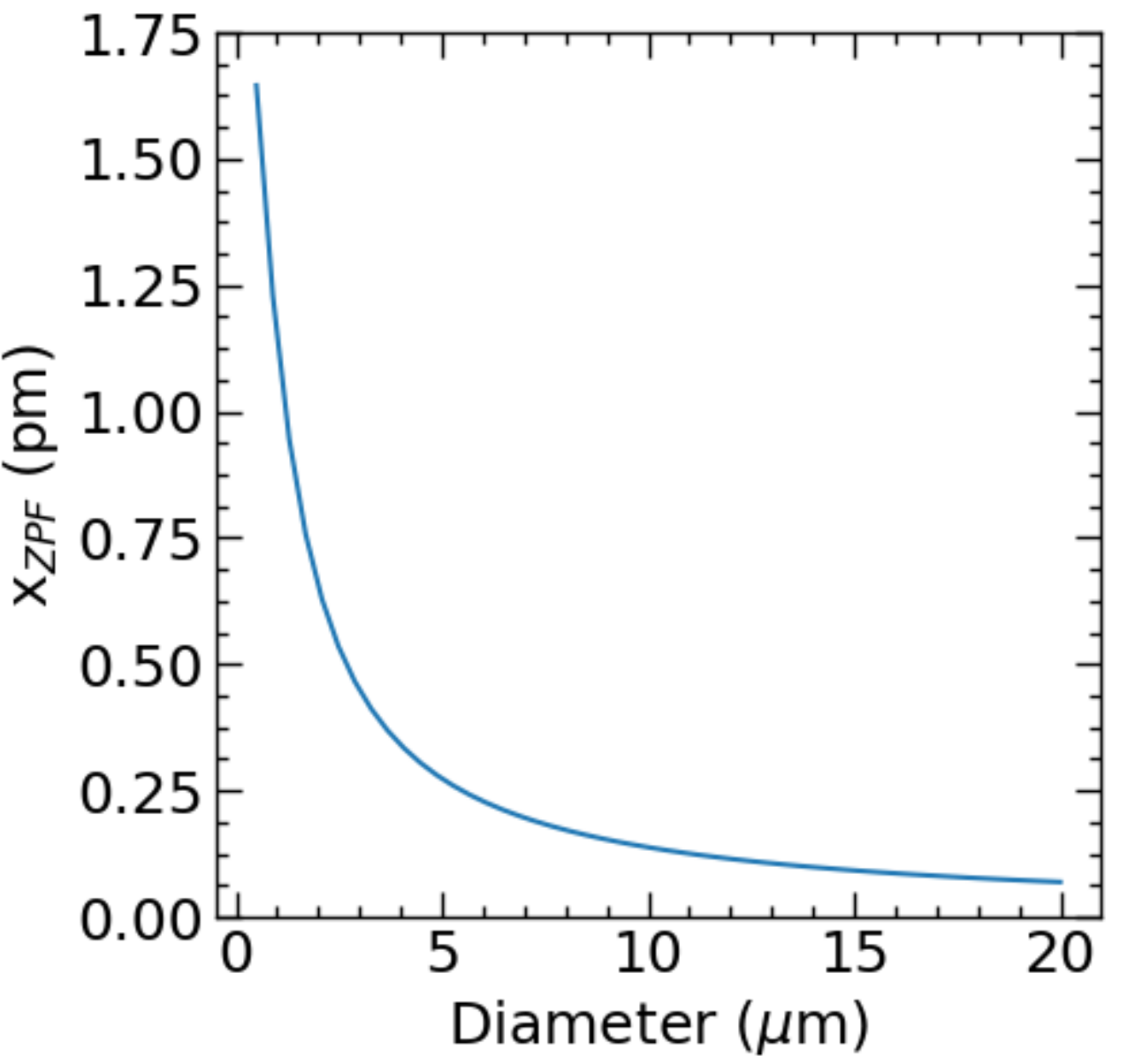}
	\caption{Zero-point-fluctuations amplitude vs microcavity pillar diameter for the $\sim 20\,$GHz confined acoustic mode.}\label{fig:xzpf}
\end{center}
\end{figure}

The optomechanical coupling rates for microcavity systems with embedded GaAs/AlAs quantum wells are determined in order to quantify the influential role of cavity polaritons. For this, the most relevant coupling mechanisms are taken into account: radiation pressure \cite{SM_Chafatinos2020} and deformation potential \cite{SM_Villafane2018}. We calculate the zero-point fluctuations associated with the vibrational coupled mode by finite elements methods, in an axial symmetry pillar, with the aim of exploiting the effect of lateral confinement \cite{SM_Villafane2018}. 
For a normalized displacement mode $\vec{u}(\vec{r})$, we can parametrize the profile as $\vec{U}(\vec{r}) = u_0 \, \vec{u}(\vec{r})$. The effective mass is obtained by the requirement that the potential energy of this parametrized oscillator is equal to the actual potential energy, and reads
\begin{equation}
\label{pot}
\frac{1}{2}\Omega_{\text{M}}^2 \int d\vec{r} \rho(\vec{r}) |\vec{U}(\vec{r})|^2 = \frac{1}{2} m_\text{eff} \Omega_\text{M}^2 u_0^2.
\end{equation}
The effective mass is rewritten as 
\begin{equation}
\label{m_eff}
m_\text{eff} = \frac{\int d\vec{r} \rho(\vec{r}) |\vec{U}(\vec{r} )|^2}{u_0^2} \equiv \int d \vec{r} \rho(\vec{r}) |\vec{u}(\vec{r})|^2,
\end{equation}
where $\rho(\vec{r})$ is the scalar density-distribution field for the structure. The reduction point for the mechanical modes $\vec{r}_0$ is chosen where the displacement is maximum ($| \vec{u}(\vec{r}_0) |= 1$). In our mechanical mode of interest the reduction point lies at the interfaces of the cavity spacer with the first sandwiching DBR layers. The spread of the mechanical coordinate in the ground state $\langle \hat{x}^2 \rangle = x_{zpf}^2$ gives the zero-point fluctuations amplitude $x_{zpf}=\sqrt{\hbar/2 m_{eff} \Omega_m}$ \cite{SM_RMP}. The effective mass related to the fundamental vibrational mode ($\sim 20\,$GHz) \cite{SM_FainsteinPRL2013} is calculated. In Fig. \ref{fig:xzpf} the zero-point fluctuations amplitude is plotted as a function of pillar diameter.

For a driven cavity, the single polariton $g_\mathrm{0}$ is amplified as $g_\mathrm{eff}=g_\mathrm{0}  \sqrt{N_\mathrm{p}}$,~\cite{SM_RMP} with $N_\mathrm{p}$ the polariton occupation. Thus, in the polaritonic condensed phase the optomechanical strong-coupling regime ($g_\mathrm{eff} > \Gamma_\mathrm{m},\Gamma_{LP}$) becomes accessible, due to the larger coherence time of the collective phase as compared to the decay time of individual polaritons. Based on the coherence time measured in our devices (in the range of 1--2~ns in condensates), with $g_\mathrm{0}/2\pi \sim 50$~MHz,  $\Gamma_\mathrm{m} \sim 50$~MHz and $\Gamma_{LP} \sim 10$~GHz, we conclude that as few as $2 \times 10^3$ polaritons in the condensate would be required to attain this condition. The ultra-strong-coupling regime, which in addition requires that $g_\mathrm{eff} > \omega_m$~\cite{SM_Forn-Diaz2019, SM_Frisk2019, SM_Hughes2021}, also becomes accessible with the studied platform. Indeed,  for  $\omega_\mathrm{m}/2\pi \sim 20$~GHz we find that  around $10^5$ polaritons would be required to attain this condition, with structures displaying optical and mechanical Q factors of $5 \times 10^4$.

\end{document}